\documentclass[11pt]{article}

\usepackage{amsmath}
\usepackage{graphicx,psfrag,epsf,ulem}
\usepackage{enumerate}
\usepackage{natbib}
\usepackage{url} 
\usepackage{amssymb}   
\usepackage{xcolor}    
\usepackage{amsthm}
\usepackage{rotating}
\usepackage{epsfig}  
\usepackage{multirow} 

\def\bSig\mathbf{\Sigma}

\usepackage{xcolor}
\usepackage{float,caption,subcaption}

\def \bx {\mbox{\boldmath $x$}}
\def \bX {\mbox{\boldmath $X$}}
\def \bY {\mbox{\boldmath $Y$}}
\def \by {\mbox{\boldmath $y$}}
\def \bbeta {\mbox{\boldmath $\beta$}}

\begin{document}
\title{A Model-free Approach for Testing Association}
\date{}
\author{Saptarshi Chatterjee\thanks{
			\textit{This project was part of Saptarshi Chatterjee's thesis completed prior to joining Eli Lilly and Company.}}\\
			Eli Lilly and Company, Indianapolis, IN,
		chatterjee\_saptarshi@lilly.com \\[10pt]
		Shrabanti Chowdhury\\
		Icahn School of Medicine at Mount Sinai, New York, NY, shrabanti.chowdhury@mssm.edu\\[10pt]
		Sanjib Basu\\
		University of Illinois at Chicago, 
		sbasu@uic.edu \\[7pt]
}
	\clearpage\maketitle
	\thispagestyle{empty}

\vspace*{-0.25in}
\begin{abstract}
	The question of association between outcome  and feature is generally framed in the context of a model on functional and distributional forms. Our motivating application is that of
	identifying serum biomarkers of angiogenesis, energy metabolism, apoptosis, and inflammation, predictive of recurrence after lung resection in node-negative  non-small cell lung cancer patients with tumor stage T2a or less. We propose an omnibus approach for testing association that is free of assumptions on functional forms and distributions and can be used as a black box method. This proposed maximal permutation test is based on the idea of thresholding, is readily implementable and is computationally efficient.
		We illustrate that the proposed omnibus tests  maintain their levels and have strong power as black box tests for detecting linear, nonlinear and quantile-based  associations, even with outlier-prone and heavy-tailed error distributions and under nonparametric setting. We additionally illustrate the use of this approach in model-free feature screening and further examine the level and power of these tests for binary outcome. We compare the performance of the proposed omnibus tests with comparator methods in our motivating application to identify preoperative serum biomarkers associated with  non-small cell lung cancer recurrence in early stage patients.
	\end{abstract}
\noindent%

	\maketitle

	\section{Introduction}
	%
	The question of association between outcome and feature  is of interest in most scientific investigations. 
    %
	This association is usually explored within the framework of an appropriate statistical model depending on the scale of the outcome, such as a linear model, a generalized linear model or a time-to-event response model. 
	%
	Nonparametric regression models provide flexibility from the constraints of specified functional forms of these parametric models. In most cases, the primary objective of such analysis  is to model the functional relationship. Instead, our objective here is to provide a decision on the association without necessarily modeling the functional relation. 
	
	Our motivation for this work comes from our longtime association with Non-Small Cell Lung Cancer (NSCLC) research. 
	Lung cancer is the leading cause of cancer mortality in the United States and NSCLC accounts for about $85\%$ of the lung cancer cases. National Comprehensive Cancer Network and Medicare and Medicaid services are supporting widespread
	implementation of lung cancer screening programs for identification of early stage lung cancers. Unfortunately,
	approximately 1 in 5 patients with pathologic stage IA
	NSCLC die of disease recurrence within 5 years of tumor resection.
	A recent study focused on identifying serum biomarkers for predicting recurrence after lung resection in node-negative NSCLC patients with tumor stage T2a or less (tumors less than 4 cm). Preoperative serum specimens of the patients were evaluated in a blinded manner for biomarkers of angiogenesis, energy metabolism, apoptosis, and inflammation; biological processes known to be associated with metastatic progression. From a statistical perspective, the popular approach for assessing association with the binary outcome of recurrence within 5 years is a binary regression analysis within the parametric framework of a logistic regression model. However, none of the biomarkers are found to be  marginally significantly associated with recurrence in logistic regression framework (see Table \ref{tab:real_data}). Penalized variable selection approaches such as Lasso \citep{tibshirani1996regression}, Elastic Net \citep{zou2005regularization} and Surely Independent Screening \citep{fan2010sure} all yield the null model as the selected model.

	Our primary objective in this article is to develop a general framework for providing a decision about the association between outcome  and feature without necessarily modeling their functional relation. Towards this goal, we propose an omnibus test for association  that can be used as a general black box tool. This omnibus test is based on 
	thresholding and is computationally efficient. 
	%
Thresholding is a popular approach to provide flexibility from or within parametric models. Recursive partitioning models (Trees \citep{breiman1984classification}, Random Forests \citep{breiman2001random}) extend thresholding to model-free approaches. Even within parametric models, thresholding provides robustness from outlying values and  assumed functional forms.

	Thresholding can be based on  scientific knowledge and historical knowledge  from previous studies. For example, Carcino Embryonic Antigen (CEA) in Table \ref{tab:real_data} is a common prognostic and predictive biomarker for NSCLC \citep{grunnet2012carcinoembryonic,dong2016serum}.
	However, there is no strong biological evidence that supports a single threshold for CEA \citep{grunnet2012carcinoembryonic}. In absence of prior scientific knowledge, thresholds are often determined empirically.
	Data dependent threshold selection approaches include splits at arbitrary percentiles (such as median).
	A popular method, as reviewed in \cite{mazumdar2000categorizing}, is an empirical systematic search for an optimal threshold which, using statistical criterion, maximizes the measure of differences between low and high groups with respect to
	an outcome. This method is associated with the minimum $p$-value approach.
	A well documented issue of this approach is the significant inflation of type I error and the resulting overstatement of  significance of the relationship between the prognostic and the outcome variable \citep{mazumdar2000categorizing}.
	In our simulation studies we found that the inflation of type-I error can be as high as $0.30$ from the target value of $0.05$.
	Several methods have been proposed for adjustment of chi-square tests under threshold selection.
	\citet{lausen1992maximally} proposed the
	maximally selected rank statistic which is the maximum of the absolute
	value of the standardized two sample linear rank statistic obtained
	through empirical process. They also derived the
	asymptotic null distribution of the statistic and showed that the
	result is similar to the asymptotic distribution of the maximum chi-square
	as derived by \citet{miller1982maximally}. Based
	on this asymptotic distribution  \citet{miller1982maximally} derived  $p$-value
	adjustment in the setting of binary outcome variables; this approach works
	best when number of potential cutpoints is large \citep{mazumdar2000categorizing}. For small minimum $p$-values, \citet{altman1994danger} provided
	a simplified formula for adjustment of the minimum $p$-value.
	\citet{lausen1996evaluating} derived a modified version of the Bonferroni
	correction which incorporates  the dependencies among the test statistics
	for adjacent thresholds. The alternative
	adjustments of the minimum $p$-value suggested by \citet{altman1994danger} and
	\citet{lausen1996evaluating} work for smaller number of thresholds. 
	
	Each of these methods has its own complexities and remains infrequently used.
	%
	Further,  we find that many of these proposed adjustments result in substantial loss of power. In this article, we instead propose a test based on the concept of permutation distribution of the maximal test statistic which can successfully capture the dependencies among the tests at individual thresholds.
	The simplicity of this method is in its generality. We illustrate that the proposed method can  be used as a general black box test for association  and does not require complex analytic adjustments. Further, being based on the permutation distribution, this method is free of distributional assumptions.
	The proposed approach is based on thresholding of the feature variable.
	There is often concern that thresholding may result in loss of information. Our simulation studies, however, show that the proposed method based on repeated thresholding of the feature variable provides comparable power under correct model specifications and superior  power under model mis-specifications compared to standard methods that utilize original feature values. 

	The rest of the article is organized as follows.  In section \ref{sec:3} we develop the general framework for testing the association hypothesis. The maximal permutation test is proposed and described in section \ref{sec:5}. Section \ref{sec:binary} develops and illustrates the utility of the test as a black box test for association in the setting of binary outcome and compares with standard and other existing methods. We illustrate the model-free  performance of the test in section \ref{sec:cont} in a wide range of settings ranging from quantile regression to heavy-tailed and outlier-prone cases. We additionally illustrate performance of the proposed approach in feature screening and compare with screening based on distance correlation \citep{Li2012}.  In section \ref{sec:6} we illustrate the performance of our proposed method in establishing association of NSCLC recurrence with preoperative serum biomarkers
	and conclude with a brief discussion in section \ref{sec:7}. 
	%
	
	\section{Formulation \label{sec:3}}
	We are interested in exploring the association between outcome $Y$ and feature $X$. The null hypothesis of no association between $Y$ and $X$ is reformulated in our thresholding approach as $\bigcap\limits_c H_{0c}$ for threshold $c$ in the feature space of $X$ and where each $H_{0c}$ represents the two-group null hypothesis of no difference between $Y|X\leq c$ and $Y|X>c$. 
	
	While our approach is completely model-free, to fix ideas, consider the case when the association between $Y$ ans $X$ is expressed by a function $h(x)$. The function $h(x)$ may be linear, 
	nonlinear, or could, for example be expressed as a nonparametric regression model.  The association with the stochastic outcome $Y$ is expressed in terms of a quantity $\eta_{Y|x}$ associated with the probability distribution $F_{Y|X=x}$ via the regression model 
	$~\eta_{Y|x} = h(x)$.
	Examples of $\eta_{Y|x}$ include $E[Y|X=x]$ or a quantile of $F_{Y|X=x}$ as in quantile regression.
	%
	Within this regression framework,
	the no association hypothesis is usually formulated as the regression function $h(x)$ being constant in $x$, namely
$	H_0: \eta_{Y|x} \equiv h_0$.
%
In the thresholding approach, we formulate this no association hypothesis as $\bigcap\limits_c H_{0c}$  where
$	H_{0c}: \eta_{Y |x\leq c} = \eta_{Y|x>c}$.
	
	To put this in a more concrete setting, popular regression models, such as generalized linear models and others, can be put in the framework where  $Y_1,\ldots,Y_n|\bX$ are independent and $Y_i|\bX=\bx \sim f(y_i|\theta_i,\phi)$ with $\theta_i = h(x_i),~i=1,\ldots,n$.
%
Within this framework, the no association null hypothesis of $H_0: h(\bx) \equiv h_0$ implies that $Y_1,\ldots,Y_n$ are i.i.d. 
Under the alternative $H_a: h(x) \neq h_0$, $Y_1,\ldots, Y_n$, however, are no longer exchangeable. 
	
We note here that one can also formulate the no association hypothesis in a nonparametric testing formulation as $\bigcap\limits_c H^{NP}_{0c}$ where 
	$H^{NP}_{0c}: F_{Y|x\leq c} = F_{Y|x>c}.$
%
	Tests focusing on $H^{NP}_{0c}$ may check for association beyond the specified regression model structure. 
	Permutation tests were introduced by \citet{fisher1936design}. The theoretical properties of these tests were studied in \citet{pitman1937significance}, \citet{lehmann1949theory}, \citet{hoeffding1952large}, among others. 
%
	Permutation tests are generally considered when the null hypothesis $H_0$ under consideration is a subset of an exchangeable specification for the outcomes $Y_1,\ldots, Y_n$,
	\begin{equation}
	H_0 \subseteq \{Y_1,\ldots,Y_n \hbox{ are i.i.d.}\} \label{eq:iid}
	\end{equation}
	
	\noindent Let $T_n(Y)=T_n(Y_1,\ldots,Y_n)$ be a test statistic for testing $H_0$. Also, let $(\pi(1),\ldots,\pi(n))$ denote a permutation of $(1,\ldots,n)$ and let $\Pi_n=\{\hbox{the group of all such permutaions } \pi \}$. The permutation distribution of the test statistic is usually constructed as $\{T_n(\pi(Y))= T_n(Y_{\pi(1)}, \ldots, Y_{\pi(n)}), ~ \pi \in \Pi_n\}$ with equal probability of $1/n!$ and the $p$-value under the permutation distribution can be roughly computed as the tail area probability of $T_n(Y)$ under this permutation distribution (The permutation test statistic is usually defined as a randomization test to account for the discreteness).
	
	Under the i.i.d. hypothesis, the permutation test is of exact level for each sample  size $n$ \citep{janssen2003bootstrap}. The book by \cite{good2013permutation} provides a practical review of permutation tests. There is an extensive literature on permutations and other resampling tests (see \citet{janssen2003bootstrap}) and studentized permutation tests when $H_0$ is strictly bigger than the i.i.d structure \citep{chung2013,JANSSEN19979}. The i.i.d. structure may be violated when the null hypothesis implies equality of a functional (such as mean or quantile) of the distributions, but not the distributions themselves. This, for example, may arise in the presence of nuisance parameters, such as, with heteroscedastic variances \citep{JANSSEN19979,chung2013}.
	
	The power of the permutation test has been extensively investigated. \citet{hoeffding1952large} established general conditions under which permutation tests are asymptotically as powerful as corresponding standard parametric tests. See also \citet{lehmann1949theory,wald1944statistical} and Lehman (1986, p230).
	
		\section{An Omnibus test for association \label{sec:5}}
	For a response $Y$ and a continuous predictor $X$, we describe here a model-free omnibus test for testing their association. The proposed test procedure can be applied when $Y$ is treated as measured in a continuous scale, or when $Y$ is binary or when $Y$ is a time-to-event, though we do not consider the last case in this article. The test procedure also does not depend on the modeled functional relation between  the predictor $X$ and response $Y$ and thus can be applied blindly irrespective of whether the functional relation is linear or nonlinear, or whether it is specified in terms of $E[Y|X|$, quantile of $F_{Y|X}$ or something else. 
	
	The concept of the test procedure is based on the observation that the hypothesis of no association typically results in $Y_1,\ldots, Y_n$ being exchangeable.
	For the test statistic, we use a repeated thresholding approach. In particular, let ${\cal C}$ denote the class of thresholds considered and for $c \in {\cal C}$, let $T_n^c(Y)$  denote the test statistic  for either the parametric hypothesis $H_{0c}: \eta_{Y|X\leq c} = \eta_{Y|X>c}$ or the nonparametric hypothesis $H^{NP}_{0c}: F_{Y|X\leq c} = F_{Y|X>c}$. We consider the test statistic
	$ T_n(Y) =  \max \{T_n^c(Y):~ c \in {\cal C}\}$
	and next apply the permutation test procedure to this test statistic $T_n(Y)$. In particular, the permutation distribution of this test statistic is given by $\{\max\limits_c T_n(\pi(Y)),~ \pi \in \Pi_n\}$
	where $\max\limits_c T_n(\pi(Y))= \max\limits_c T_n(Y_{\pi(1)}, \ldots, Y_{\pi(n)})$ for a permutation $\pi$ in permutation group $\Pi_n$.
	
	Procedurally, the proposed test is based on the following steps
	\begin{description}
	\item[$0.$] Select a two-group test statistic  $T_n^c(Y)$ for comparing $\{Y|X\leq c\}$ and $\{Y|X > c\}$.
	\item[$1.$] Obtain 
	\begin{equation}
	\max_c T_n^c(Y)= \max_c \{T_n^c(Y):~ c \in {\cal C}\}    \label{eq:maximal}
	\end{equation}
	 for the class of thresholds ${\cal C}$.
	\item[$2.$] Repeat $1.$ for $\pi(Y)$ and over the class of permutations $\pi \in \Pi_n$ to obtain the permutation distribution 
	\begin{equation}
	\{ \max_c T_n^c(\pi(Y)),~ \pi \in \Pi_n\}.\label{eq:permut}
	\end{equation}
	\end{description}
	
As is apparent from the above description, this proposed test is completely model-free and can be applied as a black box test without modeling the functional relation between Y and X. The test is also free of distributional assumptions.

The computational complexity of the test arises from the repeated computation of the test statistic over the class of thresholds ${\cal C}$ and then the repetitions over the class of permutations $\Pi_n$. However, several simplifications are possible. The indexing of $\{X \leq c\}$ and $\{X >c\}$ are invariant to permutation of $Y$ and thus can be done swiftly based on ordered $X$'s outside the repetitions over $\Pi_n$. Computational simplifications are also possible for some test statistics $T_n^c(Y)$, for example, when $T_n^C(Y)$ is chosen as the Mann-Whitney statistic, the mapping from $\bY$ to rank$(\bY)$ is invariant to permutations and thus need to be done only once for the original $\bY$. In our implementation in the R software on a personal laptop without utilizing parallel computing, for $n=400$, the whole computation for $1000$ random permutations from $\Pi_n$ was completed in 0.09 seconds.
The procedure above is, of course, ideally suited for distributing the computation of each permutation over parallel nodes.

We now discuss the choice of the two-group test statistic $T_n^c(Y)$ for comparing $\{\bY|X\leq c\}$ and $\{\bY|X>c\}$. When $Y$ is treated as measured in continuous scale, and the regression model between $Y$ and $X$ is modeled via $E(Y|X=x)$, we have $H_{0c}: E[Y|X\leq c]= E[Y|X>c]$ and  one choice for $T_n^c(Y)$ is the two-sample t-statistic based on $\{Y_i: x_i \leq c\}$ and $\{Y_i: x_i > c\}$. The choice of Welch type studentization  may provide additional robustness to the permutation test procedure (see \citet{chung2013,JANSSEN19979}) which is further discussed in section 6.1. Alternatively, and for testing $H^{NP}_{0c}: F_{Y|X\leq c} = F_{Y|X>c}$, one can use the Mann-Whitney/Wilcoxon test statistic. For binary $Y$, $H_{0c}: P(Y=1|X\leq c) = P(Y=1|X>c)$ and  $T_n^c(Y)$ can be chosen to be the chi-square or Fisher's exact statistics. For time-to-event $Y$, $T_n^c(Y)$ can be selected to be the log-rank or similar test statistics.
	
	The proposed test is of exact level by the usual theory of permutation tests as long as $Y_1,\ldots,Y_n$ are i.i.d under the no association hypothesis. The i.i.d. assumption does not hold, for example, when errors are heteroscedastic, this case is explored in section \ref{sec:6.1}. In the following sections, we examine the power of the proposed test with its comparators in different scenarios.
%
\section{Association with a binary outcome \label{sec:binary}}
	%
Consider the setting when we have observations $(\by,\bx) = \{(y_i,x_i),~i=1,\ldots,n\}$ where $y$ is treated as measured in binary scale while $x$ is continuous and the scientific question of interest is to make a decision on the association between outcome $Y$ and feature $X$. Following our approach discussed above, we reformulate the null hypothesis of $H_0$ of no association as $\bigcap\limits_c H_{0c}$ for threshold $c$ in the feature space of $X$ and where each $H_{0c}$ represents the two-group null hypothesis of no difference between $Y|X\leq c$ and $Y|X>c$.  We can construct a $2 \times 2$ table (see Table \ref{tab:1}) based on the observed $\{x_i,y_i\}$ values
	\begin{table}[h]
		\caption{Table for cutpoint $c$ \label{tab:1}}
		\begin{centering}
			\begin{tabular}{|c|c|c|}
				\hline 
				& $x_i \leq c$ & $ x_i>c$\tabularnewline
				\hline 
				$y_i=0$ & $n_{11c}$ & $n_{12c}$\tabularnewline
				\hline 
				$y_i=1$ & $n_{21c}$ & $n_{22c}$\tabularnewline
				\hline 
			\end{tabular}
			\par
		\end{centering}
	\end{table}
and let  $T_n^c(\by)$ be the $\chi^2$ test statistic based on Table \ref{tab:1} for  $H_{0c}$. Operationally, the statistic remains constant in between  ordered x-values
	$x_{(i)} \leq c < x_{(i+1)}$, so it suffices to consider the observed $\{x_i\}$ as possible choices at threshold $c$. The maximal test statistic is given as before $\max\limits_c T_n^c(\by)= \max\limits_c \{T_n^c(\by):~ c \in {\cal C}\}$
	
There is a substantial literature on the maximal $\chi^2$ test statistic based on thresholding of the feature space.
\citet{halpern1982maximally} considered the maximal $\chi^2$ statistic
over the central $\left(1-2\varepsilon\right)$ proportion of $\{x_i\}$  as   
	$\chi_{max}^{2}=\max\limits_{\left[\varepsilon n\right]+1\leq c\leq n-\left[\varepsilon n\right]-1} T_n^c(\by)$
	where $0\leq\varepsilon<\frac{1}{2}$ and $\left[K\right]$ denotes
	`the largest integer $\leq K$'. 
This method of maximal test statistics has been referred to as the equivalent minimum $p$-value approach by \citet{altman1994danger} 
to highlight the associated multiple testing \citep{mazumdar2000categorizing}. 
	%
	%
We have observed  that when $\alpha$ is controlled at $0.05$ for testing each individual $H_{0c}$ by the usual $\chi^2$ test, the family wise error	rate (FWER, \cite{Hochberg:1987:MCP:39892}) for $\bigcap H_{0c}$ can inflate to as high as $0.30$. 
	There is an extensive literature and extensive list of general approaches for controlling the FWER (see, for example, \citet{Hochberg:1987:MCP:39892,dudoit2007multiple}), however these general purpose methods may not incorporate the specific structure of repeated  thesholding. \citet{miller1982maximally} considered a similar problem, however, in their framework,  $Y=0$ and $Y=1$ groups are treated as fixed whereas $X|Y=j \sim F_j, ~ j=0,1$ and the null hypothesis of interest is $F_0=F_1$. In particular, they addressed the question of two sample comparison by the maximally selected $\chi^2$ statistic rather than the question of association between the outcome $Y$ and predictor $X$ that we are interested in. 
	A detailed review of these approaches and many other methods are discussed in \citet{mazumdar2000categorizing}.

We compare the performance of our maximal permutation test with \citet{miller1982maximally}, \citet{altman1994danger} and modified Bonferroni \citep{lausen1996evaluating} in simulation studies. For the first simulation study,  we consider a data generating model where the association between binary outcome $Y$ and predictor $x$ is  described by logistic regression 
	\begin{equation}
	logit(P(Y_i=1))=\beta_{0}+\beta_{1}x_{i},\label{logistic}
	\end{equation}
	%
For the analysis methods for evaluating the association between $X$ and $Y$, we consider $(1)$ logistic regression, $(2)$ maximal $\chi^2$ statistic based on thresholding which is then compared with $\chi^2$ distribution without any adjustments (maximal),
$(3)$ \citet{miller1982maximally}, $(4)$ \citet{altman1994danger}, $(5)$ modified Bonferroni \citep{lausen1996evaluating} and $(6)$ the proposed maximal permutation approach. We examine the level (type 1 error) and power of these approaches  by repeated data simulations from the data generating model in (\ref{logistic}).
\begin{table}[h]
 \caption{Sizes (Type-I errors) of tests. Target level is 0.05, \label{tab:type1_binary}}
	\small{
\begin{tabular}{|c|c|c|c|c|c|c|}\hline 
			 Data Generation& \multicolumn{6}{|c|}{Analysis methods} \\\hline 
          Logistic    &  Logistic & Maximal  & Miller-Siegmund & Altman  & Modified Bonferroni & Permutation \\\hline
				Linear & 0.04 & 0.32 & 0.02 & 0.02 & 0.03 & 0.04 \\
				Quadratic & 0.040 & 0.358 & 0.024 & 0.024 & 0.018 & 0.036\\\hline
			\end{tabular}}
	\end{table}

\begin{figure}[h]
		%
		%
	\begin{subfigure}[b]{0.37\textwidth}
	\centering\includegraphics[height=5cm,width=6.5cm]{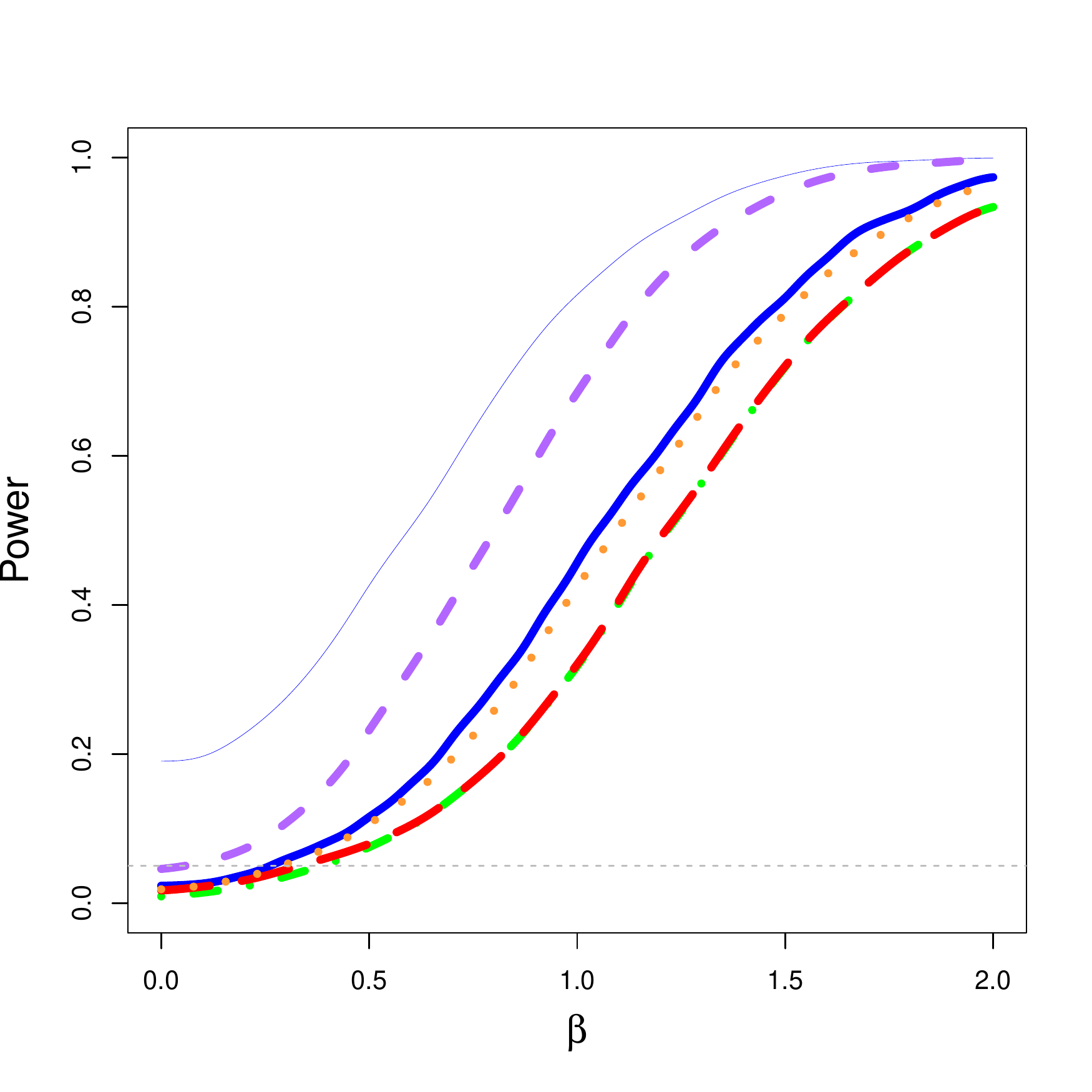}
			\caption{\label{fig:logit_correct}}
		\end{subfigure}%
		\begin{subfigure}[b]{0.37\textwidth}
			\centering\includegraphics[height=5cm,width=6.5cm]{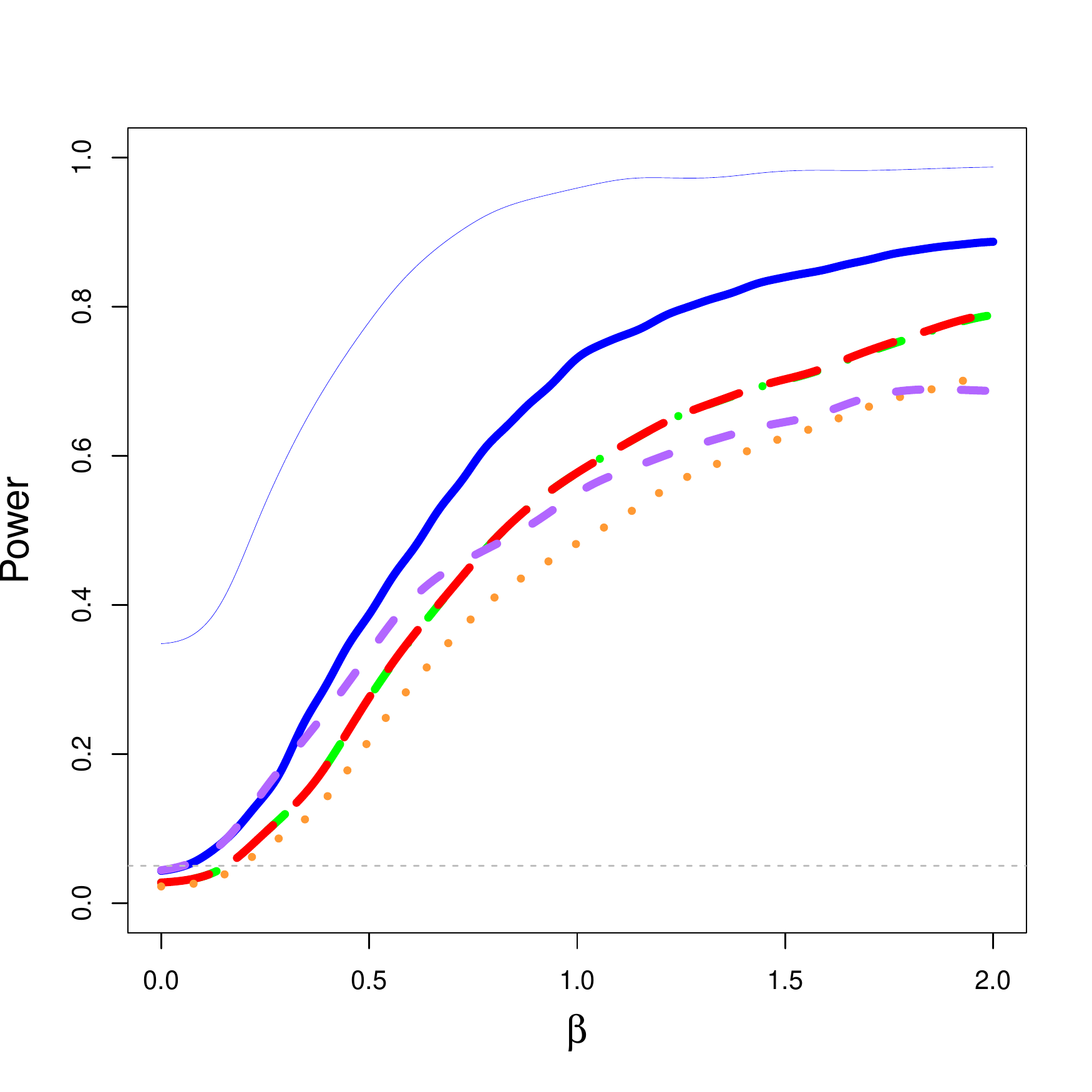}
			\caption{\label{fig:logit_misspec}}
		\end{subfigure}
		\begin{subfigure}[b]{0.22\textwidth}   
			\centering \includegraphics[height=5cm,width=4.2cm]{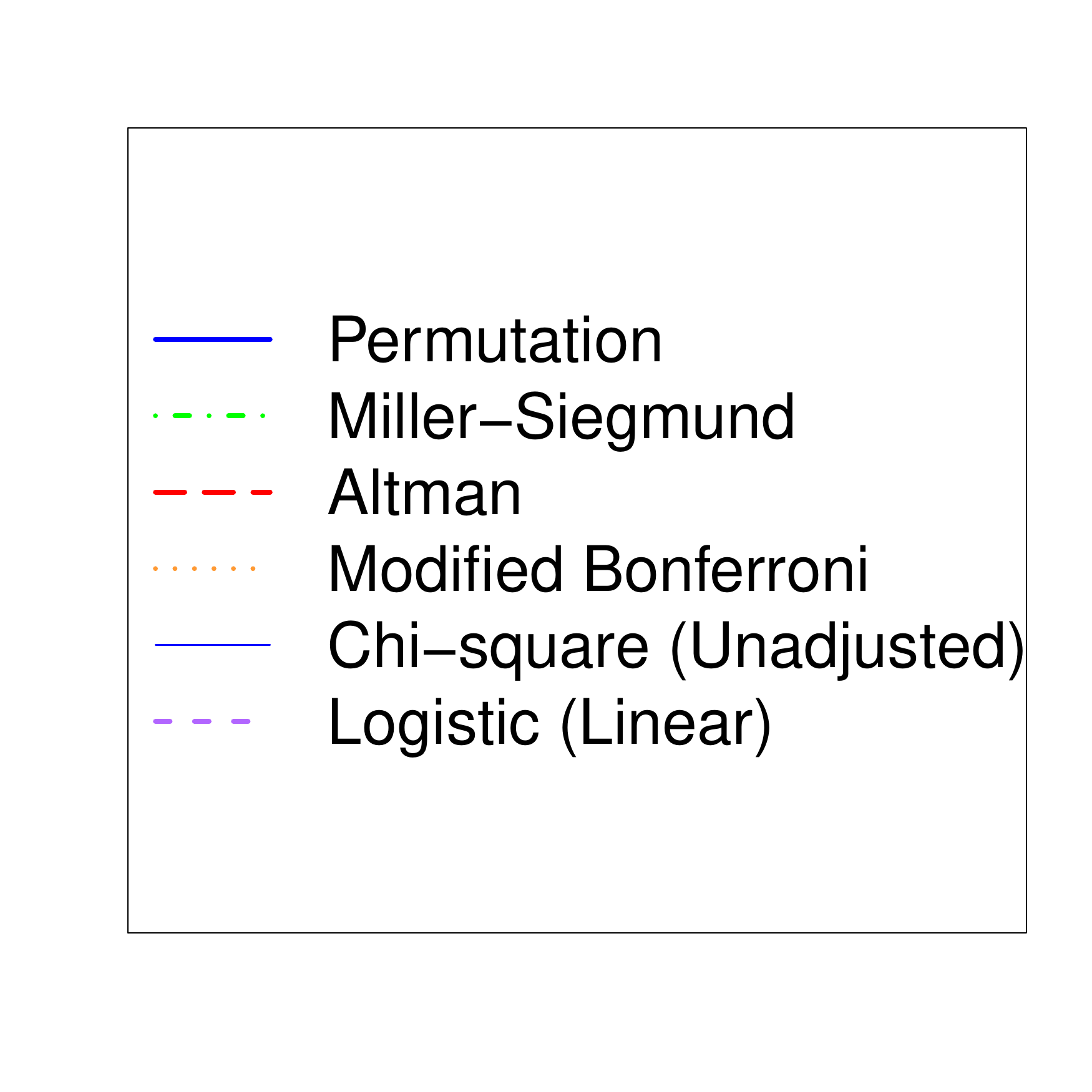}
			\caption{}    
		\end{subfigure}
		\caption{Empirically estimated power curves
		under different data generation models: (a) binary response with logistic linear regression and (b) logistic quadratic regression.} 
		\label{fig:power_binary}
	\end{figure}

Table \ref{tab:type1_binary} shows that the maximal-unadjusted approach severely inflate the Type-1 error to 0.32 from the target 0.05 level.  The previously proposed approaches of \citet{miller1982maximally}, \citet{altman1994danger} and  \citet{lausen1996evaluating}, on the other hand, are overly conservative in maintaining their levels. Figure \ref{fig:logit_correct} plots the empirically estimated  power curves of these six methods. The maximal-unadjusted approach depicts highest power but, as noted before, has substantially inflated type-I error. Among the remaining five methods which maintain their levels, the logistic regression analysis model utilizes the correct specified model here and shows highest power. The proposed maximal permutation approach displays the next best power curve. 

	%

We explore robustness to model misspecification under  with a quadratic logistic regression as the data generating model 
\[
\text{logit}\{P(Y_i=1)\}=\beta_{0} + \beta_{1}x_{i}+\beta_{1}x_{i}^2.
\]
We consider the same set of analysis methods as in the previous simulation study.
We continue to use logistic linear regression as an analysis model to mimic common practice of considering only linear relations. The maximal permutation test is used as a black box test without any modifications. The type I error reported in Table \ref{tab:type1_binary} show that the maximal unadjusted method again results in a severely inflated type 1 error. The power curves in Figure \ref{fig:logit_misspec} show that the linear-logistic method has relatively low power under this model mispecification,  The maximal permutation test shows strong power in detecting association while maintaining the level of the test whereas the previously proposed methods are conservative and have lower power compared to the maximal permutation test. 
\section{Association with outcome measured on continuous scale \label{sec:cont}}
	%
In this section, we consider the setting when we have observations $(\by, \bx) = \{(y_i,x_i),~i=1,\ldots,n\}$ where $Y$ is treated as measured on continuous scale.
The proposed maximal permutation test can again be used as a completely general blackbox method to make a decision on the association between $Y$ and $X$. We consider variants of the proposed test in (\ref{eq:permut}) based on three different choices of $T_n^c(Y)$, namely the two-sample t-statistic, the Welch t-statistic and the Mann-Whitney statistic.
	For the two-sample t-statistic,
	we take a permutation $\pi\in\Pi_{n}$ and a cutpoint
	$c\in{\cal C}$ and calculate the two-sample t-statistic 
\[
	T_{T}^{c}(\pi(Y))=\{\overline{Y}_{\pi,1}-\overline{Y}_{\pi,2}\}/\sqrt{\left\{ (n_{\pi,1}-1)\,s_{\pi,1}^{2}+(n_{\pi,2}-1)\,s_{\pi,2}^{2}\right\} /(n_{\pi,1}+n_{\pi,2}-2)}
\]
	\noindent where $(n_{\pi,1},\overline{Y}_{\pi,1},s_{\pi,1})$ and $(n_{\pi,2},\overline{Y}_{\pi,2},s_{\pi,2})$
	are the sample sizes, means and standard deviations of $\{Y_{\pi(i)}:x_{i}\leq c\}$
	and $\{Y_{\pi(i)}:x_{i}>c\}$ respectively. We next obtain $\max\limits_c T_{T}^{c}(\pi(Y))$
	and its permutation distribution over permutations $\pi\in\Pi_{n}$.
	
	There has been extensive research on two-sample permutation tests when the null hypothesis specifies equality of two population quantities (such as means) but may not result in the two population distributions being the same. One prominent example is testing for equality of means under unequal variances. \citet{JANSSEN19979} and \citet{chung2013} established that even though the exchangeable assumption does not hold in this setting, permutation test using studentized statistics, especially the Welch's t-statistic, given by 
	$T_{W}^{c}\left(\pi\left(Y\right)\right)=\left\{ \bar{Y}_{\pi,1}-\bar{Y}_{\pi,2}\right\} /\sqrt{n_{\pi,1}^{-1}s_{\pi,1}^{2}+n_{\pi,2}^{-1}s_{\pi,2}^{2}}$ asymptotically maintains the level of the test.
	
	A third alternative test statistic that we consider is the Mann-Whitney statistic which considers the nonparametric hypothesis $H_0^{NP}: F_{Y|x\leq c} = F_{Y|x>c}$ and is given by 
	$T_{MW}^{c}\left(\pi\left(Y\right)\right)=\sum\limits_{x_i\leq c} R_{pi(i)} -\{n_{\pi,1}\left(n_{\pi,1}+1\right)\}/2$
	where, $\{R_{\pi(i)}\}$ are the ranks associated with  $\{Y_{\pi(i)}\}$. Usually, rank based statistics are computationally more expensive compared to their parametric counterparts. In our setting however, the Monte Carlo permutation distribution of $\max\limits_c  T_{MW}^{c}\left(\pi\left(Y\right)\right)$ can be obtained substantially more efficiently than, for example, the t-statistics based counterparts. This substantial gain in computational efficiency is obtained from the simple observation that the mapping of rank $R_i$ to observation $Y_i$ remains invariant under a permutation, so the ranks only needs to be computed once and then reused in each permutation. 
\subsection{Linear model \label{sec:6.1}} 
	We consider a linear regression model for data generation where we treat $Y$ as measured in continuous scale and the regression
	model between outcome $Y$ and predictor $X$ is described by
	\begin{equation}
	Y_{i}=\beta_{0}+\beta_{1}\,x_{i}+\varepsilon_{i},~i=1,\ldots,n\label{eq:linreg}
	\end{equation}
	where $\varepsilon_{1},\ldots,\varepsilon_{n}\stackrel{i.i.d.}{\sim}F_{0}$.
	This results in $h(x)=\beta_{0}+\beta_{1}\,x$ and the hypothesis
	of no association is $H_{0}:\mbox{ }h(x)\equiv\beta_{0}$ or $\beta_{1}=0$.
	When $F_{0}$ is Normal, the model can also be expressed as $E[Y|x]=\eta_{Y|x}=\beta_{0}+\beta_{1}\,x$. 
	
	We examine the level (type 1 error) and power of the maximal permutation test based on the two-sample t-statistic, the Welch t-statistic and the Mann-Whitney statistic by repeated data simulations from the data generating model in (\ref{eq:linreg}) with $n=50$. As comparators, we consider the usual linear regression test for $H_0: \beta_1=0$ (denoted as LM) and a robust test using the sandwich estimator of $Var\left(\hat{\beta_{1}}\right)$
	\citep{zeileis2006object} that may provide consistent estimator
	of variance of parameter estimates even under violation of model assumptions. We also include results from the often used unadjusted maximal test statistics  methods where the maximum value of the test statistic over the set of cutpoints
	$T(Y)= \max\limits_c T^{c}(Y)$ is compared with the null distribution of the test statistic $T^{c}(Y)$
	(such as the appropriate  t-distribution for the t-statistics $T^c_T(Y)$ or $T_W^c(Y)$). As expected, the unadjusted methods result in inflation of Type-1 error. What is surprising is the level of this inflation; as we note in Table \ref{tab:type1}, the Type-1 error gets inflated to as high as $0.27$ from the nominal $0.05$ level. In contrast, the permutation tests maintain their level as prescribed by theory.
	\begin{table}[t]
		\caption{Sizes (Type-1 errors) of tests. Target level is 0.05.   \label{tab:type1}}
		\centering
	\hspace*{-4ex}
		{\small
		\begin{tabular}{c||c|c||c|c|c||c|c|c||}\hline
				& & \multicolumn{4}{|c||}{Unadjusted methods} & \multicolumn{3}{|c||}{Maximal permutation statistics}\\\cline{2-9}
			Data Generation &	LM & Sandwich & T & Welch & Mann-Whitney & T ($T_T$) & Welch ($T_W$)& Mann-Whitney $T_{MW}$ \\\hline
	Linear Regression &			0.04 & 0.06 & 0.27 & 0.27 & 0.27 & 0.05 & 0.04 & 0.05 \\
	Heteroscedastic &			0.05 & 0.08 & 0.28 & 0.28 & 0.28 & 0.05 & 0.05 & 0.05 \\
	Quadratic Reg &			0.05 & 0.06 & 0.27 & 0.28 & 0.27 & 0.05 & 0.05 & 0.04 \\
Nonparametric Reg &		0.05 & 0.06 & 0.31 & 0.32 & 0.31 & 0.06 & 0.06 & 0.06 \\

	Outliers &			0.04 & 0.03 & 0.27 & 0.31 & 0.18 & 0.05 & 0.04 & 0.04 \\\hline
	t-errors, df=1 &			0.03 & 0.03 & 0.27 & 0.36 & 0.14 & 0.05 & 0.05 & 0.05 \\
	t-errors, df=2 &			0.04 & 0.04 & 0.33 & 0.32 & 0.26 & 0.05 & 0.05 & 0.04 \\
	t-errors, df=8 &			0.04 & 0.04 & 0.33 & 0.32 & 0.31 & 0.04 & 0.05 & 0.04 \\ \hline
	Quantile Reg, p=0.25 &			0.047 & 0.046 & 0.327 & 0.368 & 0.328 & 0.053 & 0.048 & 0.051 \\
	Quantile Reg, p=0.5 &			0.042 & 0.053 & 0.297 & 0.281 & 0.296 & 0.053 & 0.046 & 0.057 \\
	Quantile Reg, p=0.75 &			0.064 & 0.069 & 0.342 & 0.412 & 0.339 & 0.062 & 0.056 & 0.053 \\ \hline
			\end{tabular}
			}
	\end{table}

	    \begin{figure}[H]
		\centering
\includegraphics[scale=0.45]{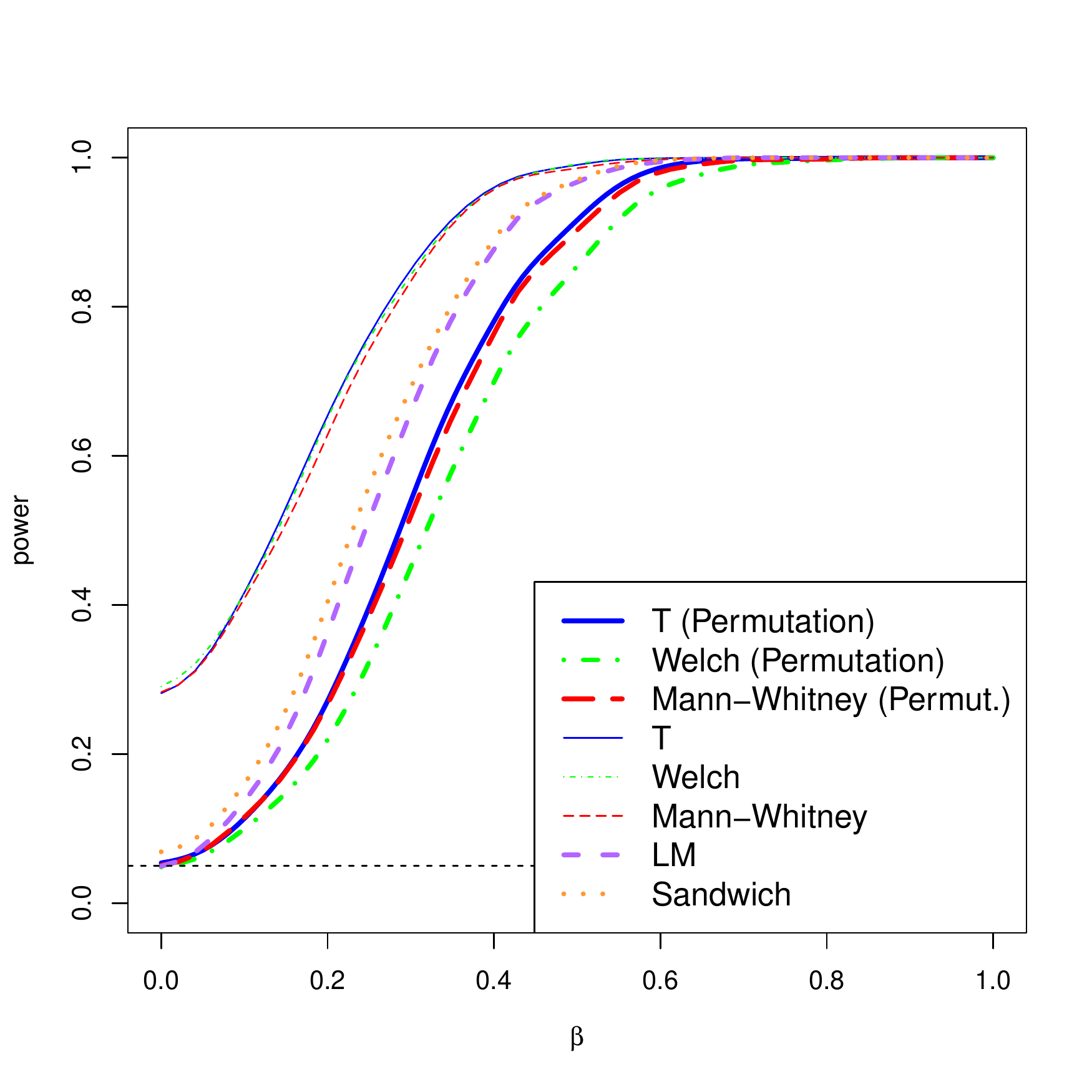}
\caption{Empirically estimated power curves under linear regression data generating model}    	\label{fig:power_curve_linreg_hom}
		\end{figure}
\begin{figure}[H]
		\centering
		\begin{subfigure}[b]{0.31\textwidth}  
			\centering
			\includegraphics[width=\textwidth]{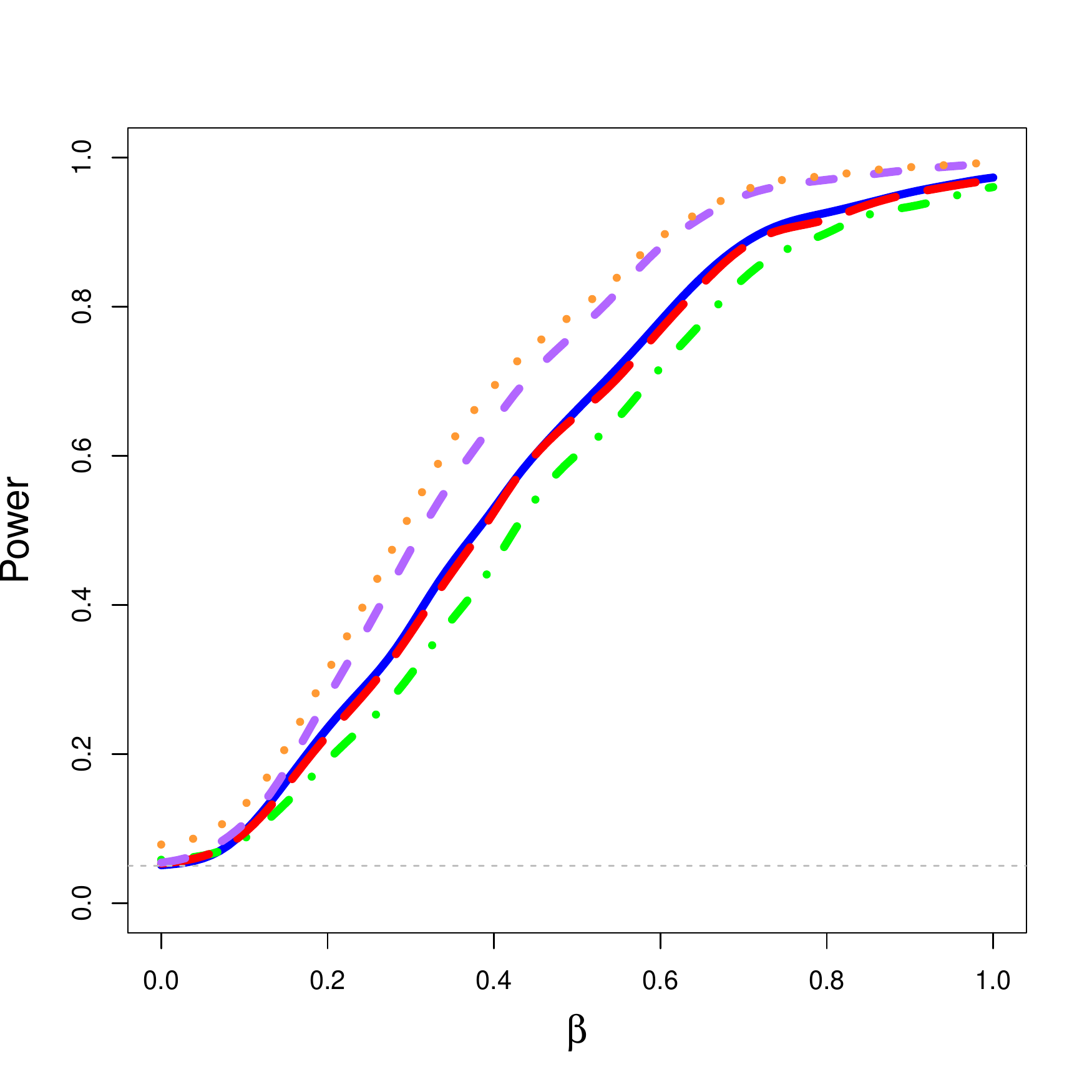}
			\caption{}    
			\label{fig:power_curve_linreg_het}
		\end{subfigure}
		\begin{subfigure}[b]{0.31\textwidth}   
			\centering 
			\includegraphics[width=\textwidth]{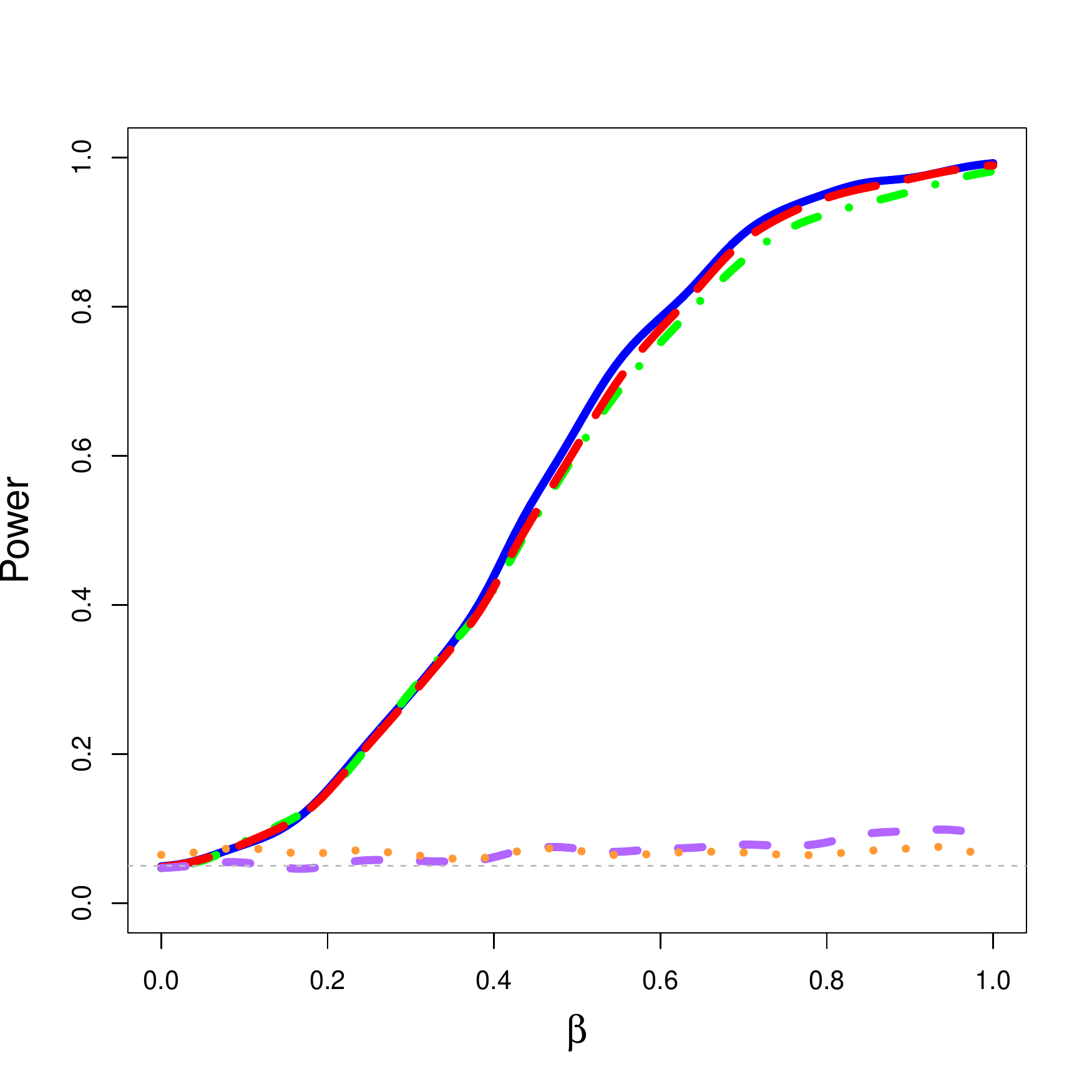}
			\caption{}    
			\label{fig:power_curve_quadreg}
		\end{subfigure}
		\begin{subfigure}[b]{0.31\textwidth}   
			\centering 
			\includegraphics[width=\textwidth]{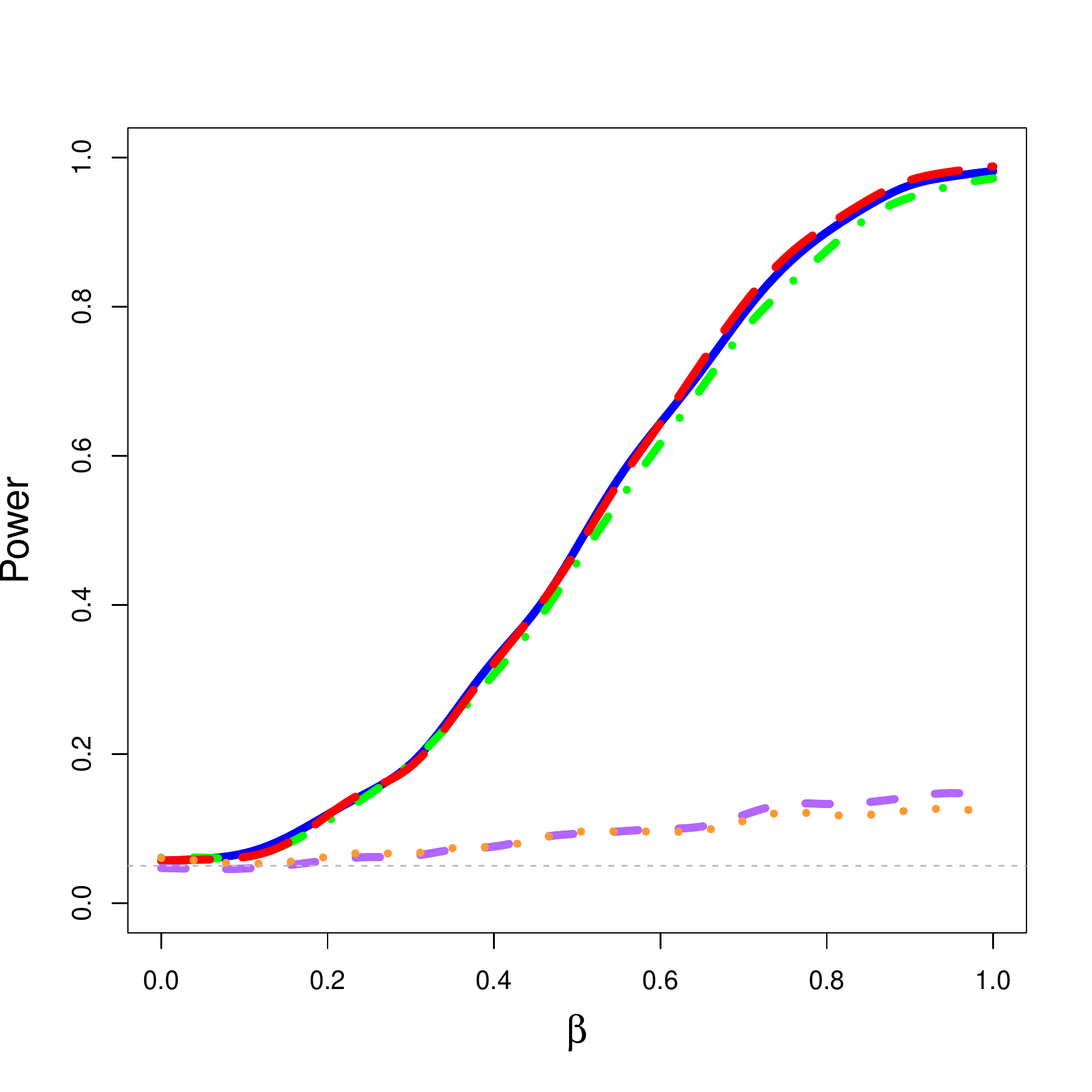}
			\caption{}    
			\label{fig:power_curve_bspline}
		\end{subfigure}
		\begin{subfigure}[b]{0.31\textwidth}   
			\centering 
			\includegraphics[width=\textwidth]{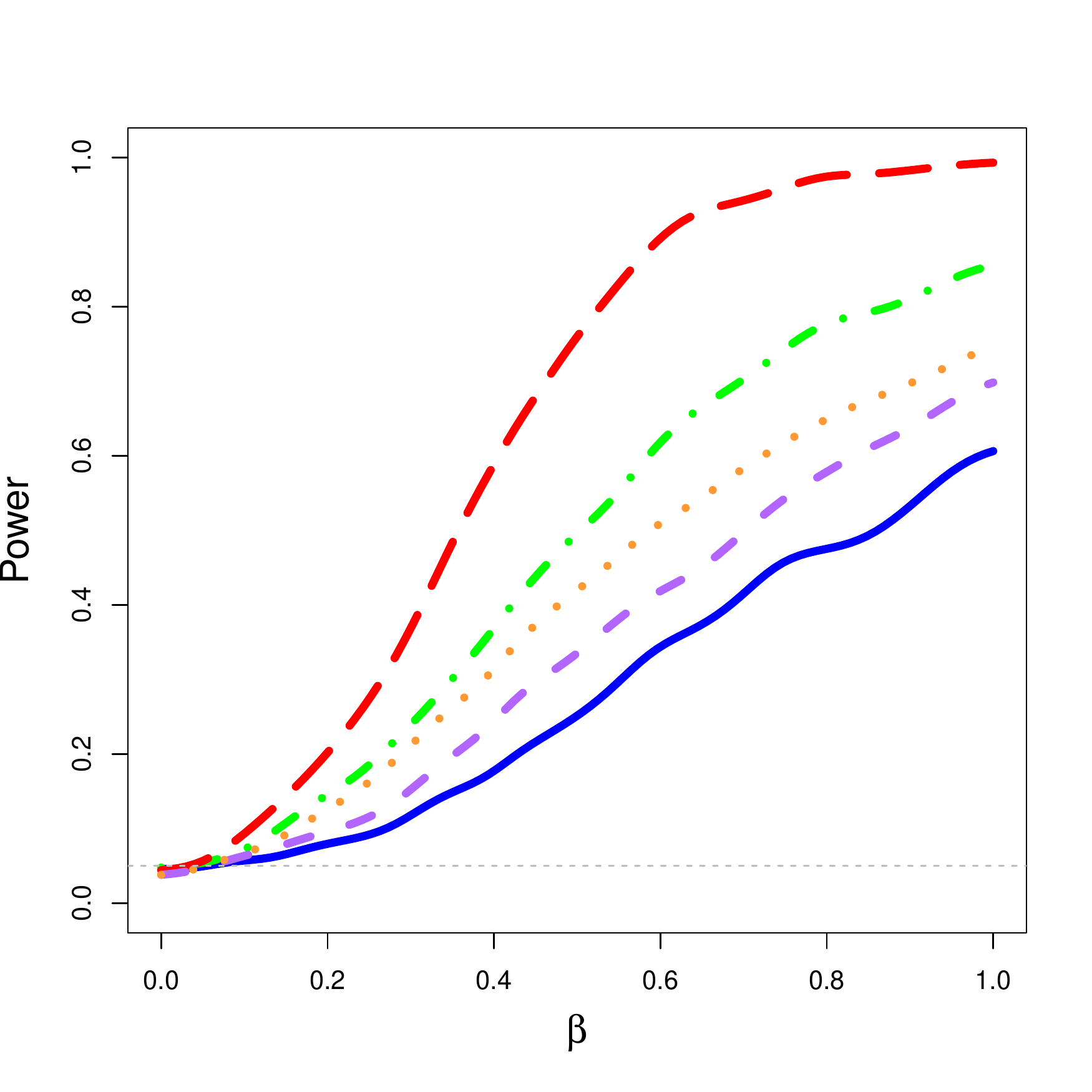}
			\caption{}    
			\label{fig:power_curve_outlier}
		\end{subfigure}
			\begin{subfigure}[b]{0.31\textwidth}
			\centering 
			\includegraphics[width=\textwidth]{legend_cont_partial}
			\caption{}    
			\label{fig:power_curve_t_df8}
		\end{subfigure}
		\caption{Empirically estimated power curves
		under different data generation models: (a) heteroscedastic errors; (b) quadratic regression; (c) nonparametric regression and (d) presence of outliers} 
	\end{figure}

Figure \ref{fig:power_curve_linreg_hom}  displays the power curves of the eight methods. The unadjusted methods have higher power curves as expected, but their power curves begin with excessively inflated levels (Type-1 error rates). The permutation based methods have only marginally lower power than the standard linear regression and sandwich methods but they provide the often sought categorization of the prognostic variable. They also provide substantial robustness (as we illustrate below) at the expense of marginal loss of power. We also note that the Mann-Whitney based permutation test maintains almost the same power as its t-test equivalent. As we noted before, the Mann-Whitney based permutation test can be computed efficiently and it is also found to be the most robust as illustrated in the following.\\[8pt] 
	\noindent \underline{Heteroscedasticity:} If the variances $Var(Y_i|x_i)$ in the linear regression model (\ref{eq:linreg}) are not equal, then even under the null hypothesis of $H_0:\beta_1=0$, $Y_1,\ldots,Y_n$ are no longer exchangeable and the basic setting under which permutation tests function do not hold. As noted before, the properties of two-sample permutation tests in this setting has been studied in \citet{JANSSEN19979} and \citet{chung2013}. We investigate a specific setting where $Var(Y|x)$ increases with increase in the value of the predictor $x$, in particular $Var(Y|x)= \sigma^2\,(1+x), x> 0$. As we notice in Table \textbf{\ref{tab:type1}}, the permutation based tests, in fact, maintain their levels even under violation of the exchangeable assumption. The estimated power curves in Figure \ref{fig:power_curve_linreg_het} shows that the permutation tests also maintain good power.\\[8pt]
	%
%
%
\begin{figure}[bt]
		\centering
\includegraphics[scale=0.2]{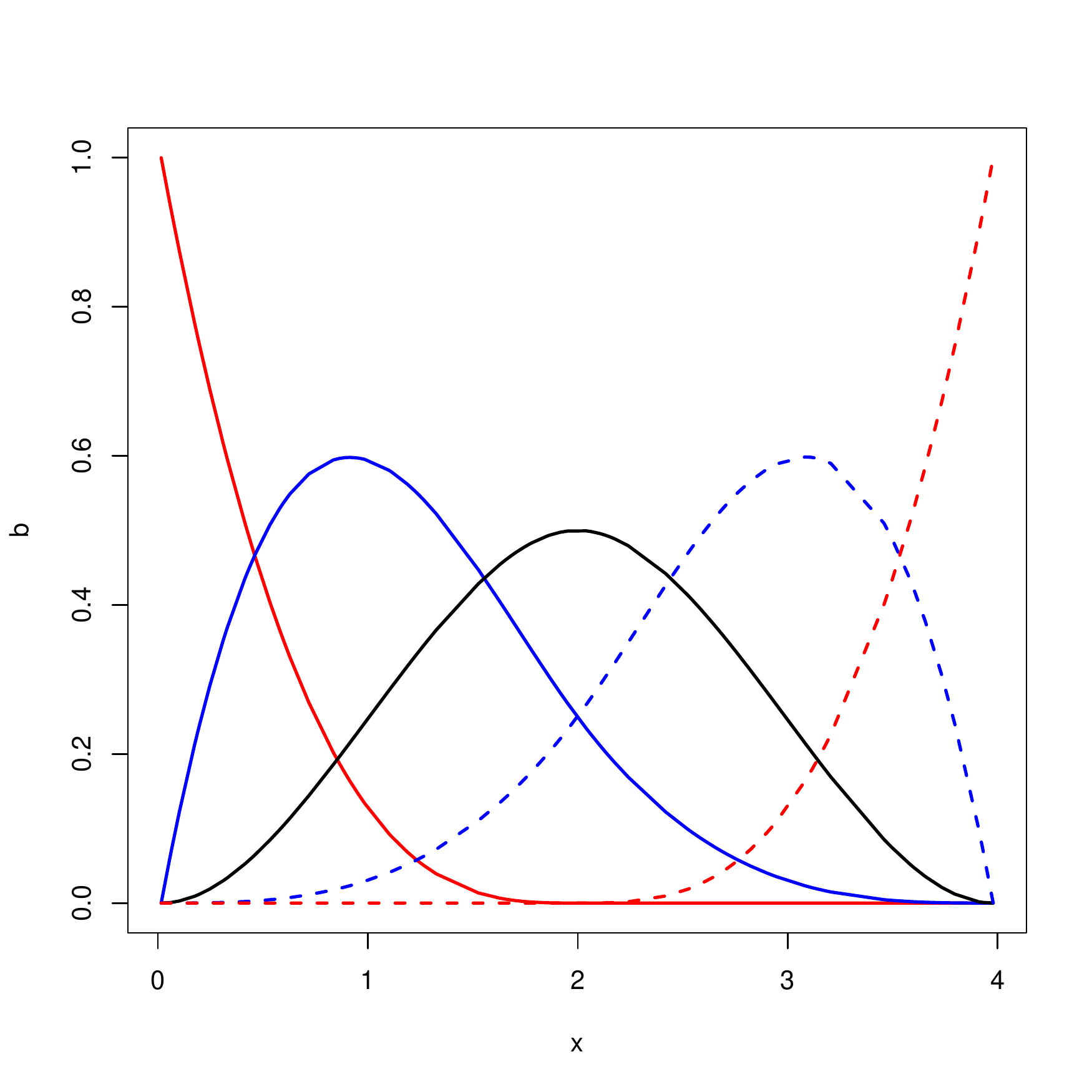}
\caption{B-spline basis functions}
			\label{fig:basis}
	\end{figure}
	\noindent\underline{Quadratic regression:} We explore the wide scale applicability of the proposed test  in the setting of a quadratic regression data generating model 
	\[
	Y_{i}=\beta_{0}+\beta_{1}\left(x_{i}-2\right)^{2}+\epsilon_{i},\mbox{ }i=1,\ldots,n
	\]
	
	In common practice of statistical analysis, only linear functional forms are explored in analysis models to test for association and we mimic this case by considering the linear functional form based tests (LM and sandwich) as tests for association. In addition we investigate the unadjusted and permutation based maximal test statistics. Table \ref{tab:type1} shows that the unadjusted tests severely inflate the Type-1 error as before. The power curve estimates in Figure \ref{fig:power_curve_quadreg} illustrate that due to the misspecification of the functional form, the linear functional form based tests (LM and sandwich) have almost no power at all in detecting the association. The maximal test statistics based permutation tests, on the other hand, display strong power for detecting association while maintaining the level of the tests. We want to emphasize that the permutation tests were used completely as generic black box tests  without any modifications to reflect the structure of the  data generating model.\\[8pt]
\noindent\underline{Nonparametric Regression:} We also explore the applicability of the proposed approach in the setting of a nonparametric
	regression model
	\[
	Y_i = f(x_i) + \epsilon_i,~ i = 1,\ldots,n
	\]
for a function $f(\cdot)$ defined on $x$-space. We model the function $f(\cdot)$ using a basis expansion as $f(\cdot) = \sum \beta_j\, f_{j}(\cdot)$ where $\{f_j(\cdot)\}$ are the basis functions.

 For our numerical study, we consider the cubic B-spline basis; the first few B-spline basis functions are shown in Figure \ref{fig:basis}. We vary the  $\beta_j$ coefficients to consider a range of increasing signal to noise ratios in the nonparametric regression model and generate replicated datasets from each of these settings. As before, we apply the tests as blackbox association tests without any input about the data generating model. The power curve estimates in Figure  \ref{fig:power_curve_bspline} illustrate that the linear functional form based tests  (LM and sandwich) again suffer from model misspecification and yield almost zero power. The maximal permutation tests, on the other hand, provide sufficient power in this nonparametric regression setting while having comparable type-I error (Table 3). 
 
	%

	\subsection{Outliers}
	%
	The robustness of the proposed tests in presence of outliers is investigated by considering a contaminated distribution \citep{tukey1960survey} for data generation from the model in (\ref{eq:linreg})  where $Var(Y|x)$ is taken to be $=1$ with probability $0.9$ and $=100$ with probability $0.1$; the large variance case allowing the potential of outliers. As in the previous cases, the permutation tests as well as the LM and sandwich tests maintain level $\leq 0.05$ whereas the levels of the unadjusted tests are severely inflated. In Figure \ref{fig:power_curve_outlier}, we note that the permutation adjusted rank based maximally selected Mann-Whitney test statistic outperforms the other methods. This is expected as rank based methods are resistant to outliers. We also note that the robust sandwich method outperforms the LM method. 
	%
	%
	\subsection{Heavy tailed error distributions}
\begin{figure}[t]
		\centering
		\begin{subfigure}[b]{0.24\textwidth}  
			\centering 
			\includegraphics[width=\textwidth]{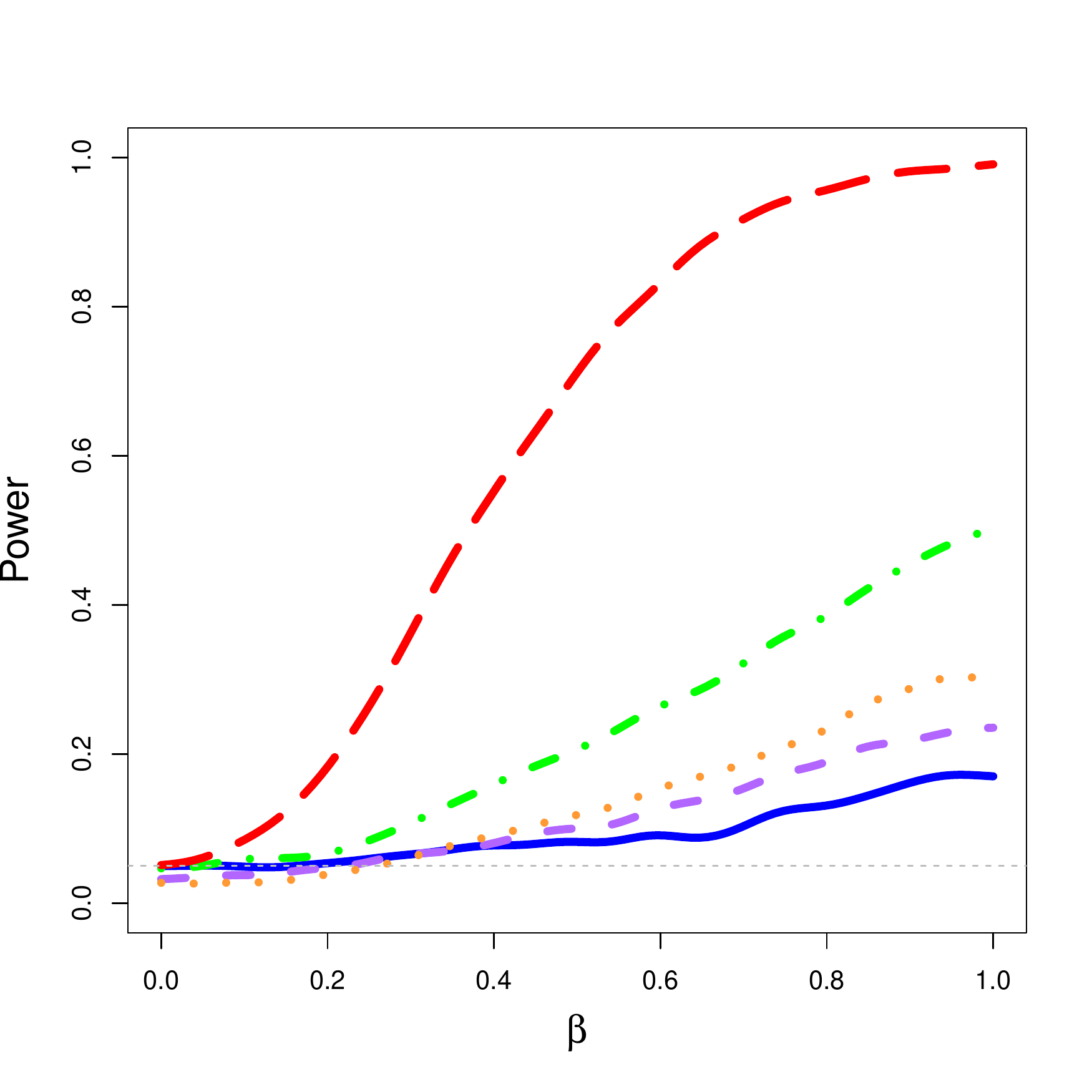}
			\caption{}    
			\label{fig:power_curve_t_df1}
		\end{subfigure}
		\begin{subfigure}[b]{0.24\textwidth}   
			\centering 
			\includegraphics[width=\textwidth]{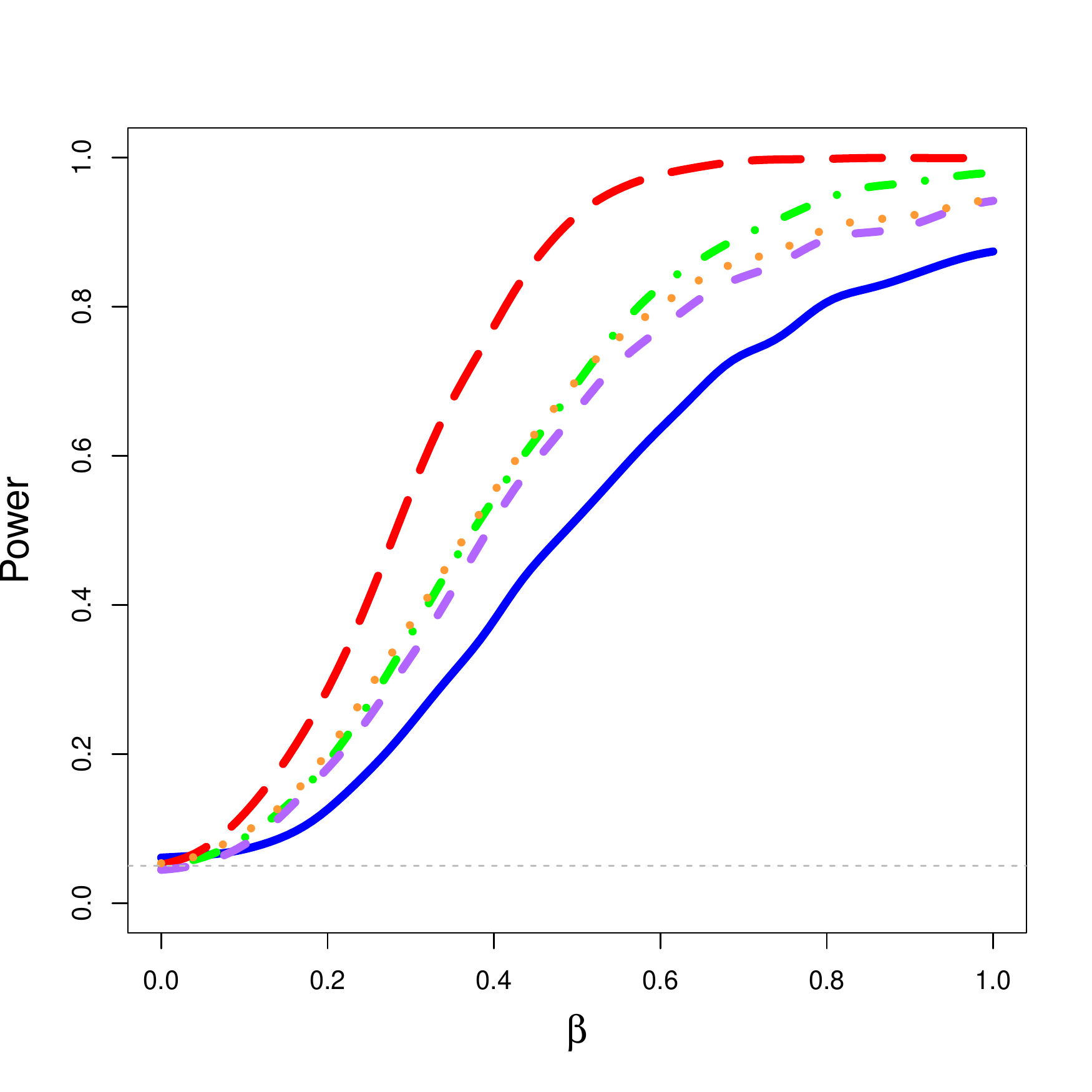}
			\caption{}    
			\label{fig:power_curve_t_df2}
		\end{subfigure}
		\begin{subfigure}[b]{0.24\textwidth}   
			\centering 
			\includegraphics[width=\textwidth]{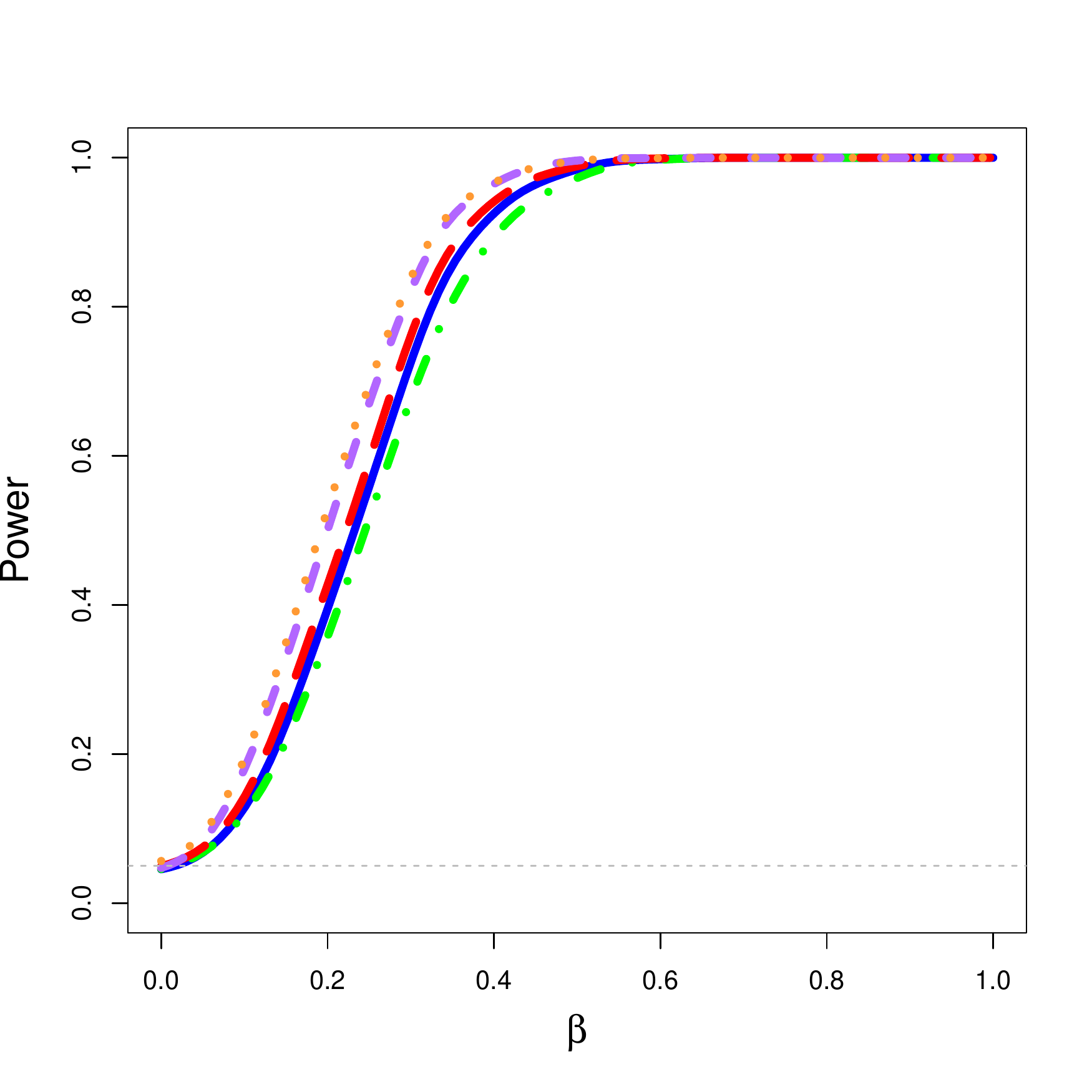}
			\caption{}    
			\label{fig:power_curve_t_df8}
		\end{subfigure}
		\begin{subfigure}[b]{0.24\textwidth}   
			\centering 
			\includegraphics[width=\textwidth]{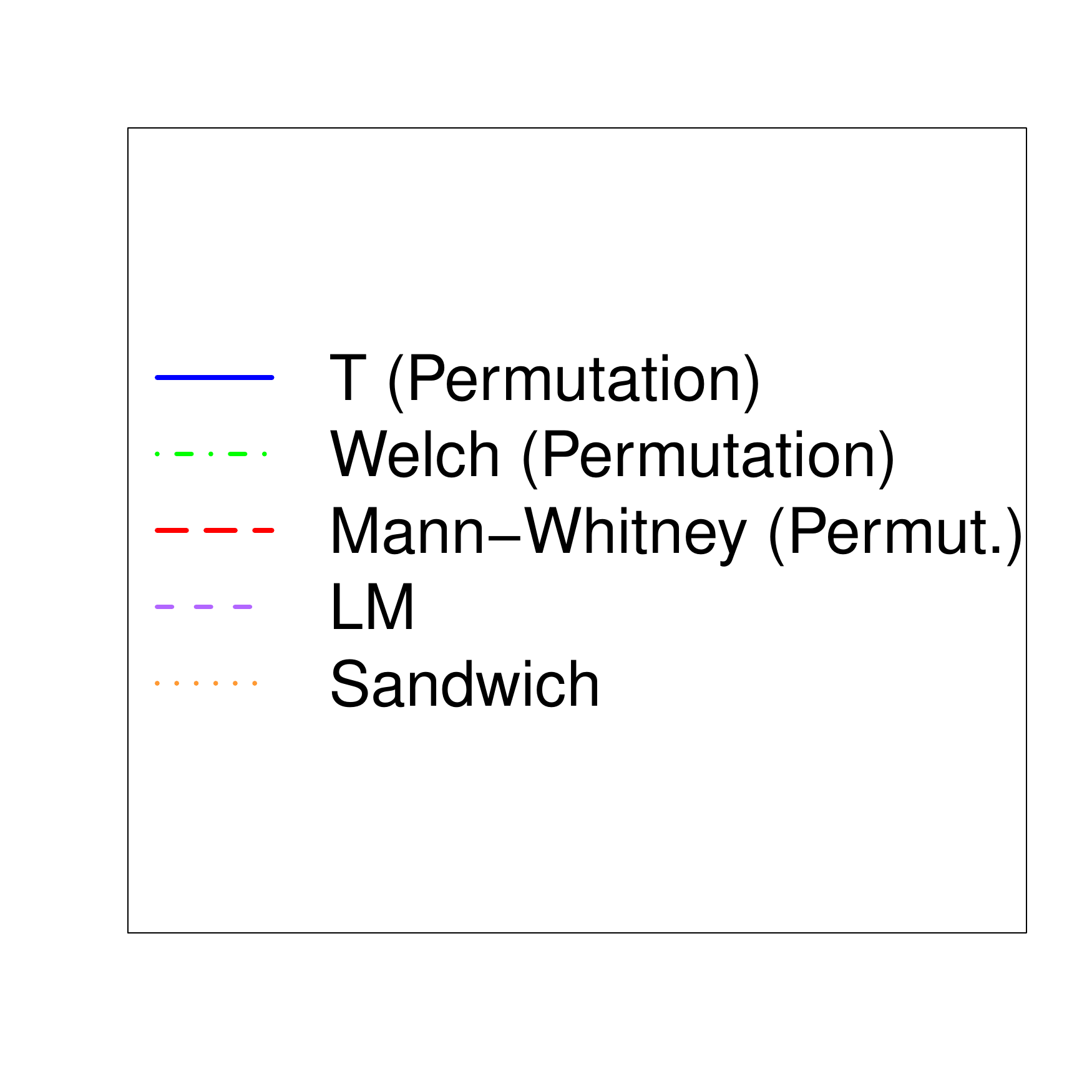}
			\caption{}    
		\end{subfigure}
		\caption{Empirically estimated power curves
		under different data generation models: (a) linear regression with errors following t-distribution, df=1; (b) t-distribution, df=2; (c) t-distribution, df=8.} 
		\label{fig:power_t}
	\end{figure}

	The robustness of the proposed tests is further investigated by considering heavy tailed error distributions in data generation from the model in (\ref{eq:linreg}). We considered t-distributions with 1, 2 and 8 degrees of freedom (df) respectively for the error distribution. Note that for $df=1$, $E(Y|x)$ does not exist and the regression model can only be specified in terms of median $(Y|x)$. As seen in Figure \ref{fig:power_curve_t_df1}-\ref{fig:power_curve_t_df8}, the LM and sandwich methods as well as t-test based permutation methods perform poorly in this case whereas permutation adjusted rank based maximally selected Mann-Whitney test statistic significantly outperforms all other methods. For t-distribution with $df=2$, $Var(y|x)$ does not exist. The  LM, sandwich and t-test based permutation methods perform better here but the permutation adjusted Mann-Whitney test statistic  continues to outperform  other methods. For $df=8$, the estimated power curves reflect an ordering similar to the nominal case ($df=\infty$) where the permutation test based methods have power slightly below the LM and sandwich methods.
	%

	\subsection{Quantile Regression}
\begin{figure}[t]
		\centering
		\begin{subfigure}[b]{0.24\textwidth}
			\centering
			\includegraphics[width=\textwidth]{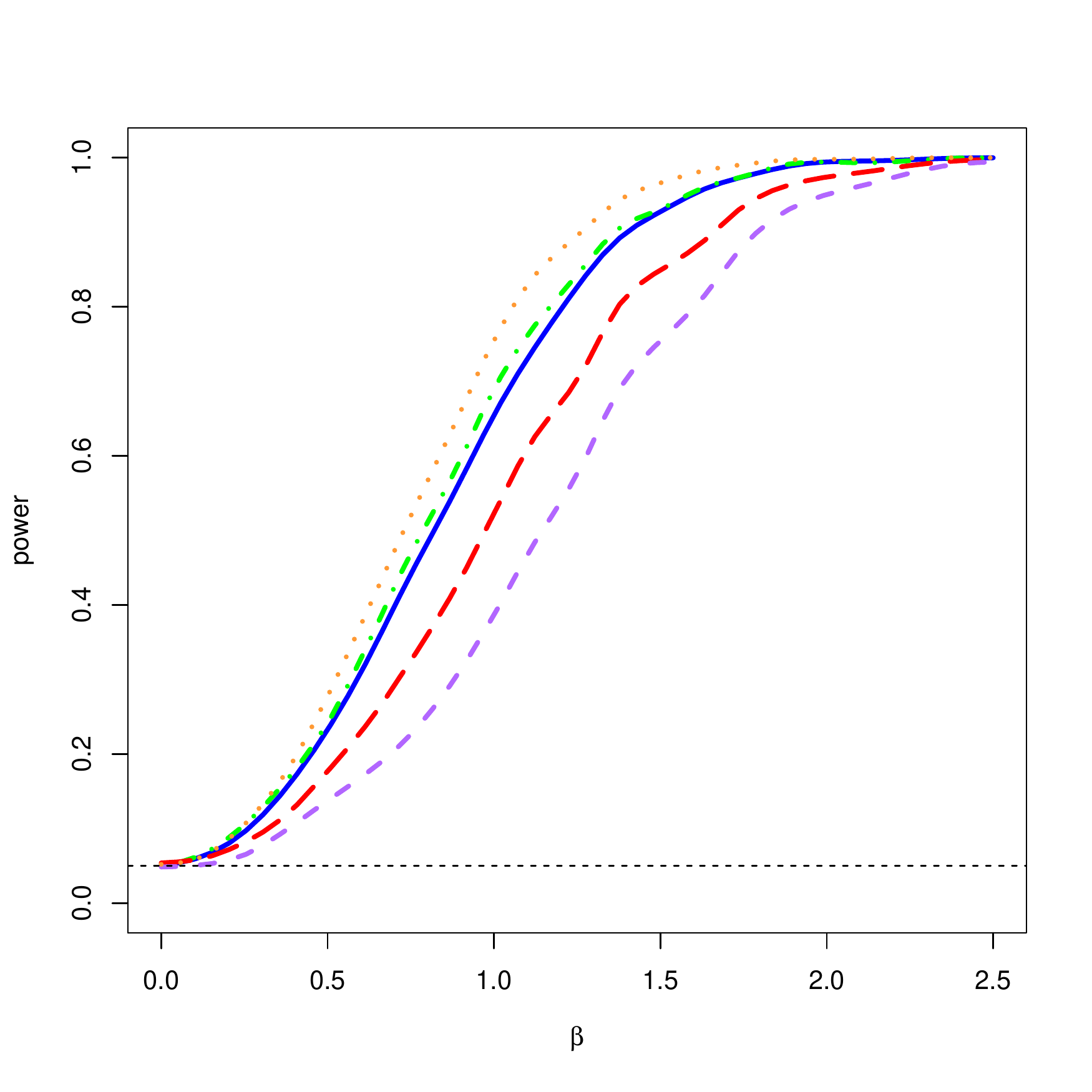}
			\caption{}    
			\label{fig:power_curve_quant_0_25}
		\end{subfigure}
		\begin{subfigure}[b]{0.24\textwidth}  
			\centering 
			\includegraphics[width=\textwidth]{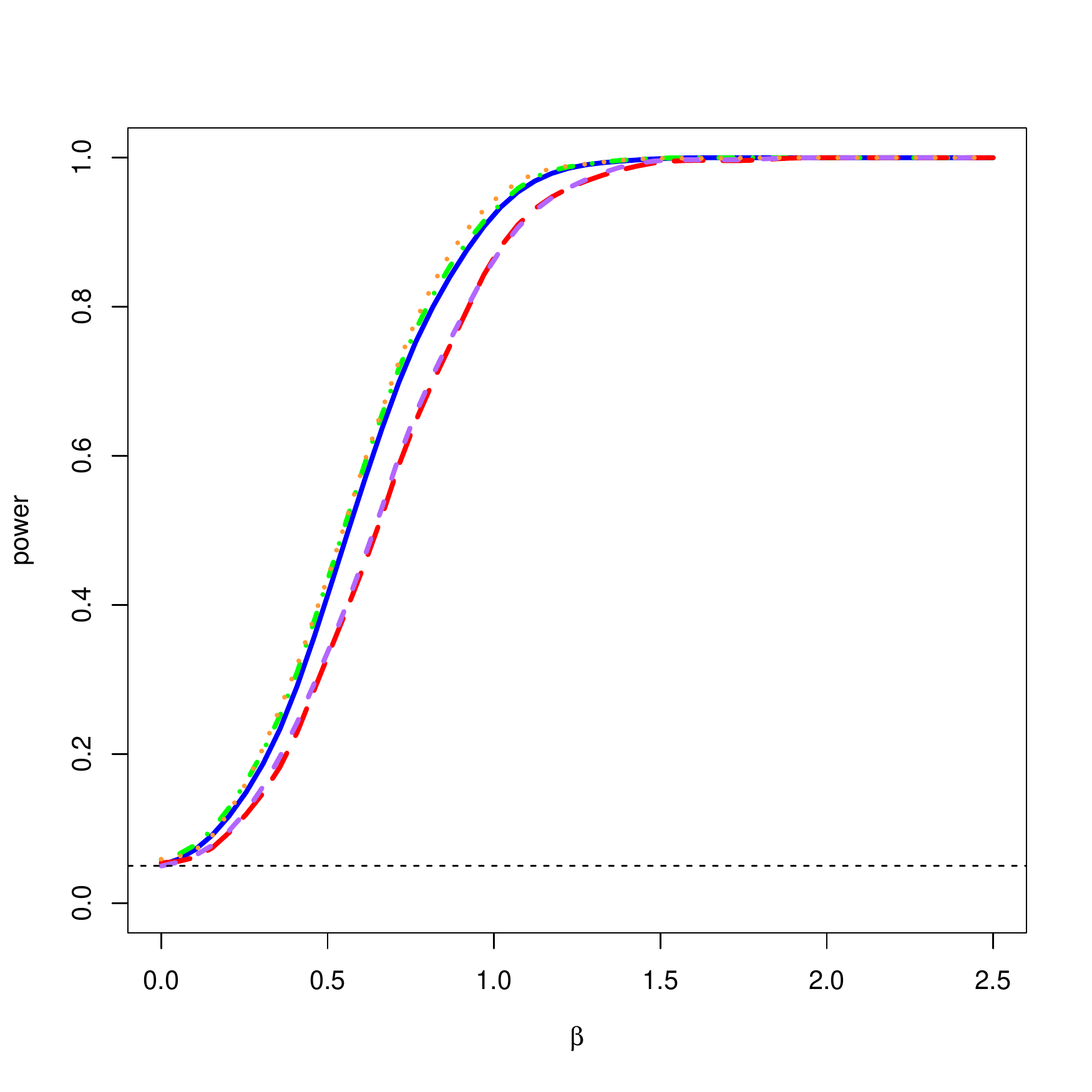}
			\caption{}    
			\label{fig:power_curve_quant_0_5}
		\end{subfigure}
		\begin{subfigure}[b]{0.24\textwidth}   
			\centering 
			\includegraphics[width=\textwidth]{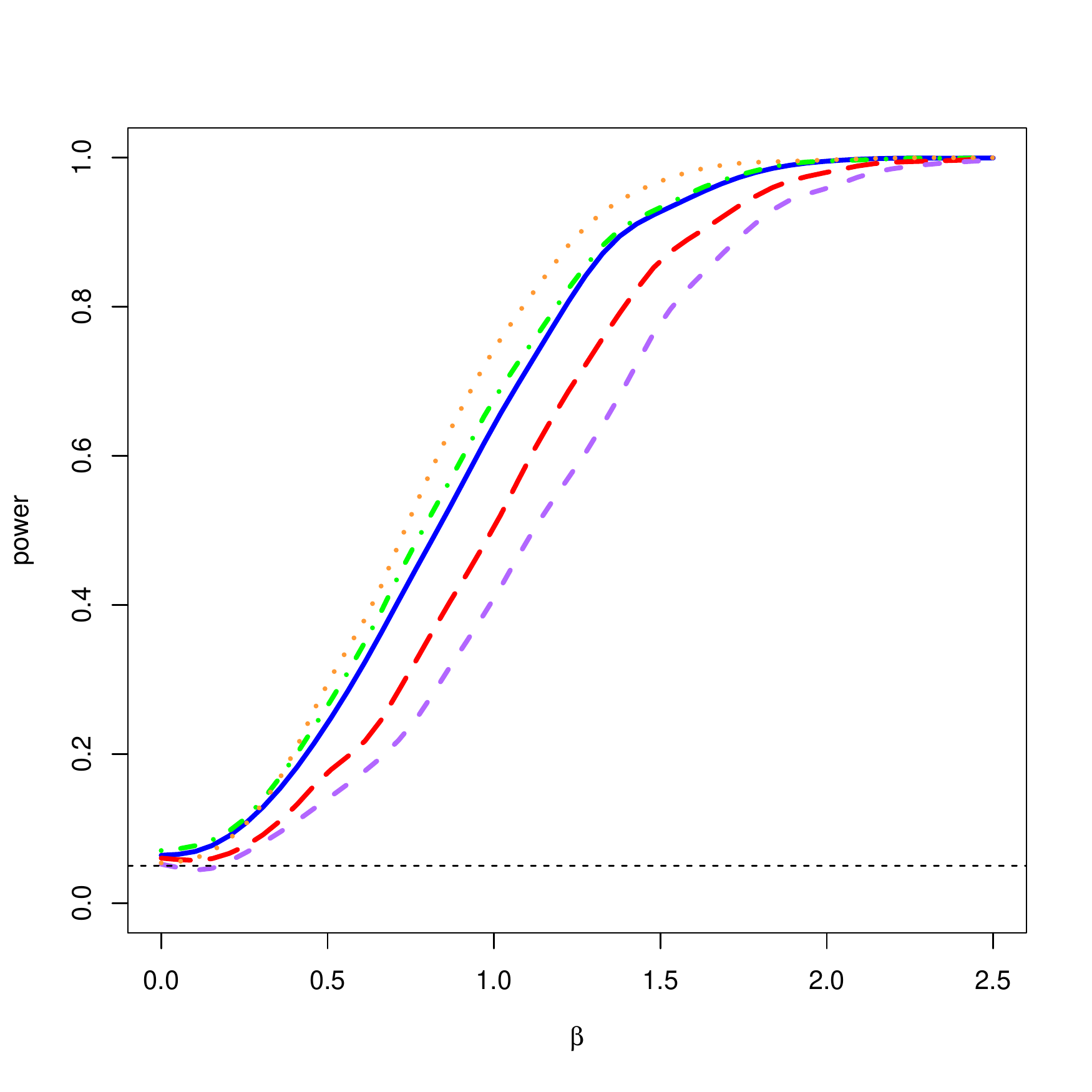}
			\caption{}    
			\label{fig:power_curve_quant_0_75}
		\end{subfigure}
				\begin{subfigure}[b]{0.24\textwidth}   
			\centering 
			\includegraphics[width=\textwidth]{legend_cont_partial.pdf}
			\caption{}    
		\end{subfigure}

		\caption{Empirically estimated power curves
		under different data generation models:  (a) quantile regression with 0.25$^{th}$ quantile;  (b) quantile regression with 0.50$^{th}$ quantile; (c) quantile regression with 0.75$^{th}$ quantile.} 
		\label{fig:power_different}
	\end{figure}

	We also investigated the robustness of our proposed test methodology in exploring the association between the outcome and predictors in the framework of quantile regression \citep{yu2001bayesian}. The $p^{th}$ quantile $\left(0<p<1\right)$ of $Y$ conditional on $\boldsymbol{X}$, denoted by $q_{p}\left(Y|\boldsymbol{X}\right)$, is regressed on a set of predictors $\boldsymbol{X}$ 
	\begin{equation}
	q_{p}\left(Y|\boldsymbol{X}\right)=\boldsymbol{X}\boldsymbol{\beta}_{p},
	\end{equation}
	where $\boldsymbol{\beta}_{p}$ is the set of regression coefficients corresponding to the $p^{th}$ quantile of the outcome. The estimate of $\boldsymbol{\beta}_{p}$, $\hat{\boldsymbol{\beta}}_{p}$, is the solution to the minimization of the loss function
	\begin{equation}
	\min_{\boldsymbol{\beta}_{p}}\sum_{t}\rho_{p}\left(y_{t}-x_{t}^{\prime}\boldsymbol{\beta}_{p}\right),
	\label{eq:qreg}
	\end{equation}  
	where $\rho_{p}\left(u\right)=u\left(p-I\left(u<p\right)\right)$. The loss function in (\ref{eq:qreg}) is considered to be robust compared to the quadratic loss function in linear regression. The quantile regression problem can equivalently be formulated in terms of the asymmetric Laplace distribution \citep{yu2001bayesian} and in our data generation model, we generate $Y_i \sim$ Asymmetric Laplace $\left(\mu_{i},\sigma\right)$ with
	$g\left(\mu_{i}\right)=\boldsymbol{x}_{i}^{\prime}\boldsymbol{\beta}_{p}$.
	We generate $X$ from $Uniform\left(0,4\right)$, consider values of  $\beta$ in $\left[0,2.5\right]$ and further considered $p=0.25,0.5\mbox{ and }0.75$ quantiles. In Figure \ref{fig:power_curve_quant_0_25}-\ref{fig:power_curve_quant_0_75}, we observe that for $p=0.25$ the permutation adjusted rank based Mann-Whitney test outperforms the rest of the methods giving maximum power while simultaneously maintaining the estimated type-I error at 0.05. 
    However, the power of the permutation adjusted two sample T test and Welch test are observed to be less than that of the LM and Sandwich test. Similar results are obtained for $p=0.75$ too. Also, the ordering of the performance of the methods remains unchanged for $p=0.5$; permutation based Mann-Whitney still outperforms other methods. 
    
\subsection{Feature screening \label{sec:filter}}
In  recent years, there has been increased interest  in feature screening or filtering while modeling association with a number of predictor variables. Screening is often used to reduce the dimensionality of the feature space so that it is amenable for the next step of the analysis.
Screening approaches in the setting of linear models include the sure independent screening (SIS) method proposed in  \cite{fan2008sure} and the forward regression in \cite{Wang2009}. \cite{Fan2009} proposed screening in generalized linear models whereas 
\citet{fan2011nonparametric} considered feature screening in nonlinear additive models.

%
%
 Consider the case when we have a number of $X$-variable $X_1, \ldots , X_p$ but the data-generating model is postulated to be sparse in which only a few of these $X$-variables are postulated to be associated with outcome $Y$.  Sparsity is a frequently assumed framework in high dimensional data.  Screening in this context is often done using a marginal utility or a marginal measure of association. In particular, let $T_j = T(Y,X_j)$ denote a measure of  marginal association between $Y$ and variable $X_j$. Features $\{X_j\}_{j=1}^p$ are then screened based on $\{T_j\}_{j=1}^p$ values, and for a pre-specified integer $k$, a screening approach proposes the reduced set of $X$ variables
\[
	 \mathcal{X}_R = \{1\leq j\leq p: T_j \mbox{ is among the top $k$ values}\}
	 \]
for consideration in the next stage of the anaysis.
Alternative strategy for constructing $\mathcal{X}_R$ may consist of including variables which are within a pre-specified threshold.
 
 \cite{fan2008sure} considered screening with $T(Y,X_j)$ being the marginal correlation between outcome $Y$ and feature $X_j$ and established sure independent screening property in the setting of linear regression. In the setting of generalized linear models, \cite{Fan2009} proposed independence screening with maximum marginal likelihood
estimators. \citet{fan2011nonparametric} considered the context of nonparametric additive model where screening is performed by fitting marginal non-parametric regression to each of the features and then thresholding the utility of the predictors.
 
In this numerical study, we propose feature screening using the proposed maximal permutation test statistic as the marginal screening statistic $T(Y,X_j)$. As we have argued, the maximal permutation statistic is model-free and can be used as a blackbox approach for testing association. We have also illustrated  strong performance of this statistic in detecting linear and nonlinear associations.
 
 We explore screening performance in an example considered in  \citet{meier2009high} and \cite{fan2011nonparametric}. In this setting of a sparse additive model, out of features $\{1,\ldots,p\}$, the sparse data generating model only includes predictors ${\cal D} = \{j_1, \ldots,j_d
\}$ which are associated with outcome $Y$ via an additive model
\[ Y_i = \sum_{j \in {\cal D}} g_{j}(x_{ij}) \; + \; \varepsilon_i,~ i=1,\ldots,n\]
where $g_j(\cdot)$ are functions of the respective predictors. Following \citet{meier2009high} and \cite{fan2011nonparametric}, we take $d=4$, ${\cal D}= \{1,2,3,4\}$ and $g_1(x) = \beta_1\,x$,~ $g_2(x) = \beta_2\,(2x-1)^2$,~ $g_3(x) = \beta_3\, \frac{ \sin(2\pi x)}{(2-\sin(2\pi x))}$ and $g_4(x) = \beta_4 \{\,0.1\sin(2\pi x) \,+\, 0.2\cos(2\pi x) \,+\,
    0.3 \sin(2\pi x)^2 \,+\, 0.4 \cos(2\pi x)^3 \,+\,
    0.5 \sin(2\pi x)^3 \}$. 
We consider the case of $p=100$, where the remaining $96$ $X$-variables do not contribute in the data generating model.

\cite{Li2012} proposed use of the distance correlation for general feature screening  that may not require a model specification. The distance covariance \citep{distcorr2007, distcorr2009} between two random vectors is a weighted $L^2$-distance between the joint characteristic function  and the product of the marginal characteristic functions. The distance correlation is the ratio of the  distance covariance to the product of the distance standard deviations. It is a measure of dependence between the random vectors and equals zero if and only if the random vectors are independent \citep{distcorr2007}. Following the distance correlation based screening in \cite{Li2012} and other works, we consider $T(Y,X_j)$ = distance correlation$(Y,X_j)$ as a comparator screening method.

We follow the simulation framework in \cite{meier2009high} and \cite{fan2011nonparametric} and consider $n=400,~p=100$. $\varepsilon_i \stackrel{i.i.d.}{\sim} N(0$, variance=1.74), and $\bbeta =(\beta_1,\ldots,\beta_4) = (5,3,4,6)$. We additionally include $\bbeta=(0,0,0,0),~ (1,1,1,1)$ and $(2.5,1.5,2,3)$ to consider cases of no and weak associations.  We compute  $\{T_l(Y,X_j),~ j=1,\ldots,p,~ l=1,2\}$ where $T_1(\cdot)$ and $T_2(\cdot)$ are respectively the maximal permutation test statistic and the distance correlation. For each $T_l$, we perform marginal screening  by keeping only those $X_j$ having the top $k$  $T_l(Y,X_j)$ values. We consider three choices, $k=4, 10$ and $20$. 
The results in Table \ref{tab:screen} are based  on 100 replications.

For the null case of $\bbeta=(0,0,0,0)$, the chance of an individual feature $j$ being among the top $k$ is completely random and we notice that the empirical inclusion probabilities from permutation screening closely align with these values. For the non-null cases, screening performances improve as signal-to-noise increases in $\bbeta$ ranging among $(1,1,1,1)$, $(2.5, 1.5, 2,3)$ and $(5,3,4,6)$. We notice that associations with predictors 1 and 2 are easier to detect by screening whereas that of predictor 3 is more difficult to screen.  In  general, Table \ref{tab:screen} illustrates that the maximal permutation based screening performs comparatively, and often in a superior fashion in including relevant predictors in the screening set. We also note that computation time for maximal permutation based screening was in the same scale as that of distance correlation based screening. 
\begin{table}[t]
    \centering
\caption{Empirical inclusion probabilities of individual predictors and all predictors in the data generating model after screening}
\label{tab:screen}
\scriptsize
\hspace*{-2ex}
    \begin{tabular}{|cc|ccccc|ccccc|ccccc|ccccc|}\hline
 \multicolumn{2}{|c|}{} &\multicolumn{5}{|c|}{$\bbeta=(0,0,0,0)$} &\multicolumn{5}{|c|}{$\bbeta=(1,1,1,1)$}
 &\multicolumn{5}{|c|}{$\bbeta=(2.5,1.5,2,3)$} &
 \multicolumn{5}{|c|}{$\bbeta=(5,3,4,6)$}\\\hline
 \multicolumn{2}{|c|}{Screening} & 1 & 2 & 3 & 4 & All 4 & 1 & 2 & 3 & 4 & All 4 & 1 & 2 & 3 & 4 & All 4 & 1 & 2 & 3 & 4 & All 4 \\\hline
   Top 4  & Permutation & 
   .02 &  .02 &  .01 & .04 & .00 &
   1 & .97 & .56 & .45 & .22 &
   1 & .97 & .63 & .99 & .60 &
   1 & .98 & .76 & 1 & .75 \\
   & Dist. Corr. &
  .07 & .07 & .05 & .04 & .00 &
  1 & .96 & .39 & .30 & .09 &
  1 & .96 & .49 & .91 & .36 &
  1 & .92 & .59 & .98 & .52 \\\hline
 Top 10 & Permutation &
  .11 & .10 & .10 & .16 & .00 &
  1 & 1 & .85 & .80 & .66 &
  1 & 1 & .89 & 1 & 0.88 &
  1 & 1 & .94 & 1 & .94\\
   & Dist. Corr. &
.12 & .13 & .09 &  .12 & .00 &
1 & 1 & .73 & .56 & .43 &
1 & 1 & .82 & .99 & .81 &
1 & .98 & .91 & 1 & .90 \\\hline
 Top 20 & Permutation &
 .17 & .19 & .17 & .27 & .00 &
 1 & 1 & .89 & .93 & .82 &
 1 & 1 & .95 & 1 & .95 &
 1 & 1 & 1 & 1 & 1 \\
   & Dist. Corr. &
.23 & .21 & .20 & .24 & .00 &
1 & 1 &  .89 & .79 & .70 &
1 & 1 & .92 & 1 & 0.92 &
1 & 1 & 1 & 1 & 1 \\\hline
    \end{tabular}
\end{table}

	\section{NSCLC recurrence and baseline biomarker levels \label{sec:6}}
	
    A primary motivation for the statistical methodological development in this article is based on our work in NSCLC research. Lung cancer is  the leading cause of cancer deaths in the United States, with approximately 222,500 new diagnoses and 157,600 deaths estimated in 2017. A large proportion of patients who undergo lung resection for non-small cell lung cancer dies of disease recurrence within 5 years. Numerous observational studies have shown low-dose computed tomography (LDCT) of the lung to be an effective method for screening lung cancer, especially in the early stage in high-risk patient population \citep{doria2010screening}. The National Comprehensive Cancer Network recently recommended such screening for appropriately selected high-risk patients \citep{national2011reduced} and the Centers for Medicare
	and Medicaid Services have approved payment screening in these patients \citep{mulshine2014issues}.
	Nevertheless, there is a concern over the accuracy of lung cancer screening as it involves moderately high number of false positives and consequent morbidity. Unfortunately, approximately 1 in 5 patients with pathologic stage IA NSCLC die of disease recurrence within 5 years of tumor resection. Blood and serum-based biomarkers, including EarlyCDT-lung and microRNA based biomarkers, serve as potentially useful supplements to LDCT for lung cancer screening, by further evaluating patient risk prior to LDCT, or assessing malignant risk of positive LDCT findings \citep{kanodra2015screening}. The identification of prognostic biomarkers of NSCLC patients is crucial from both clinical and therapeutic decision perspectives. Clinical and demographic variables such as male gender, age, non-squamous histology, are known to have negative prognostic effects for NSCLC patients \citep{williams1981survival,charloux1997prognostic}. We consider data from a recent NSCLC study on $n=123$  early stage (Stage 1A and 1B) patients who underwent lung resection and the objective is to identify serum biomarkers predictive of recurrence after lung resection. Preoperative serum specimens of the patients are evaluated in a blinded manner for biomarkers of angiogenesis, energy metabolism, apoptosis, and inflammation; biological processes known to be associated with metastatic progression. Biomarker levels were measured using the Luminex system. Recurrence was considered as that occurring within 5 years from the date of surgery. The median follow up was 58.2 months and  23 patients had NSCLC recurrences during follow-up. 
	
\begin{figure}[ht]
		\centering
		\begin{subfigure}[b]{0.33\linewidth}
			\centering\includegraphics[height=4cm,width=4cm]{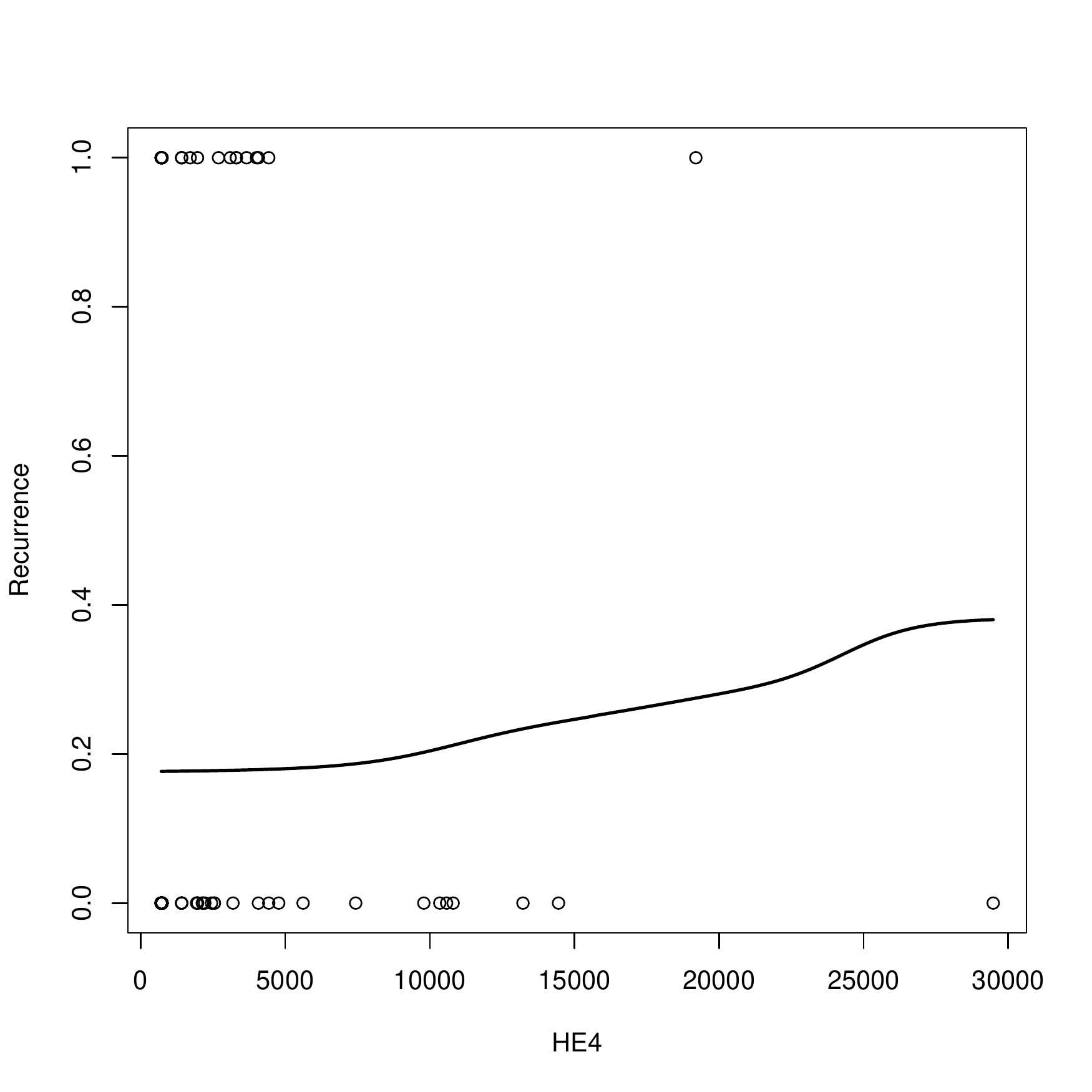}
			\caption{\label{fig:HE4_scatter}}
		\end{subfigure}%
		\begin{subfigure}[b]{0.33\linewidth}
			\centering\includegraphics[height=4cm,width=4cm]{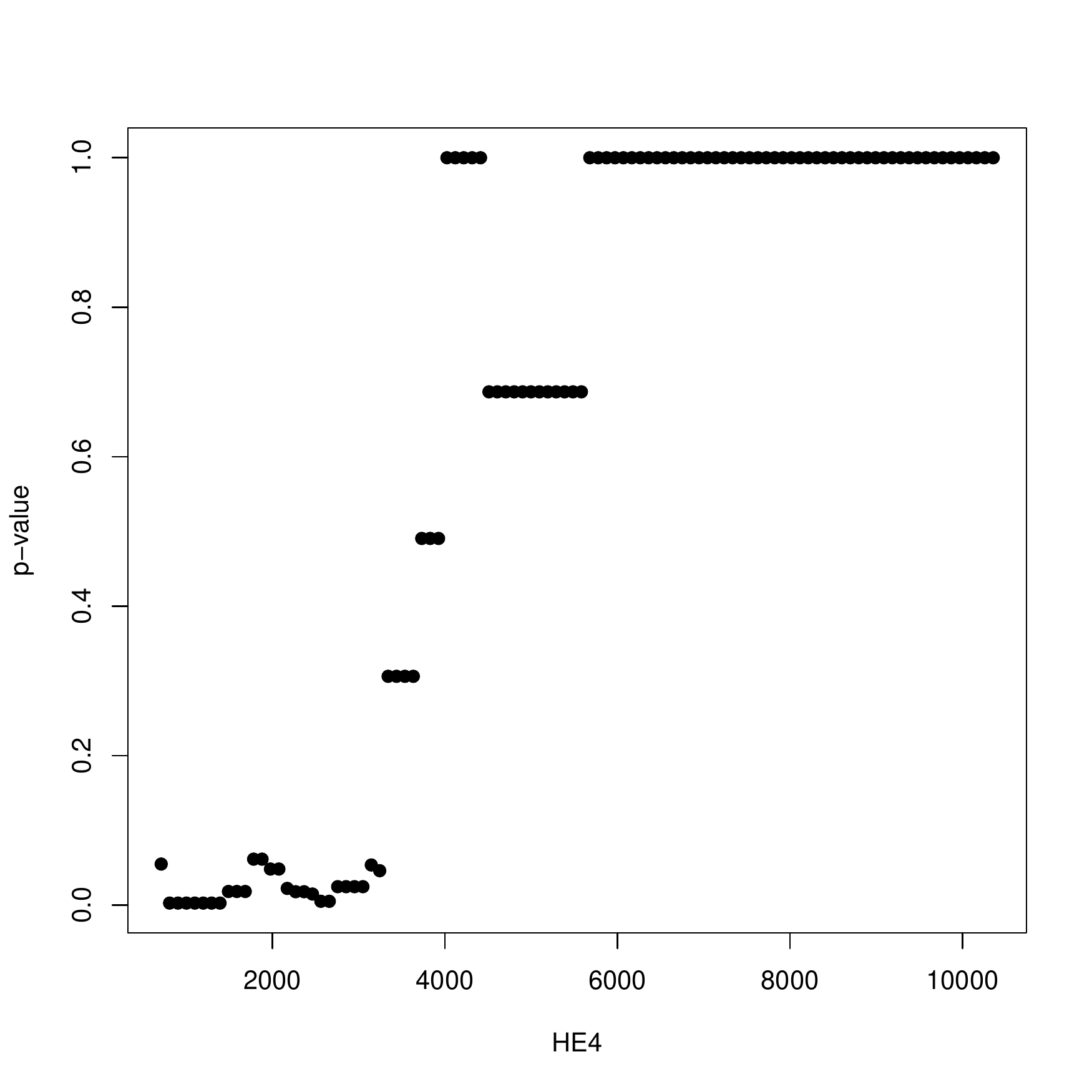}
			\caption{\label{fig:HE4_pvalplot}}
		\end{subfigure}\\
			\begin{subfigure}[b]{0.24\textwidth}
			\centering
			\includegraphics[width=\textwidth]{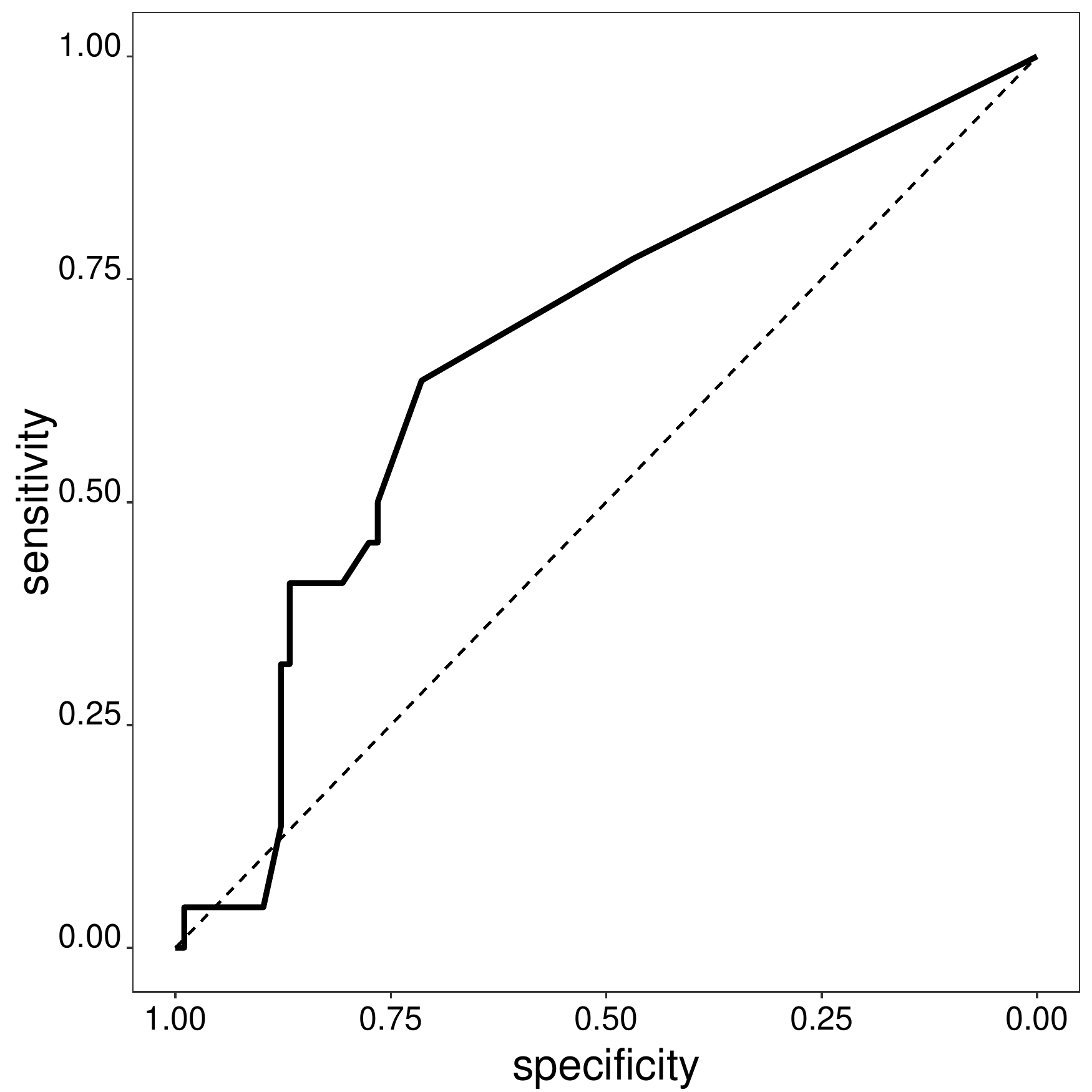}
			\caption{HE4}    
			\label{fig:roc-HE4}
		\end{subfigure}
		\begin{subfigure}[b]{0.24\textwidth}  
			\centering 
			\includegraphics[width=\textwidth]{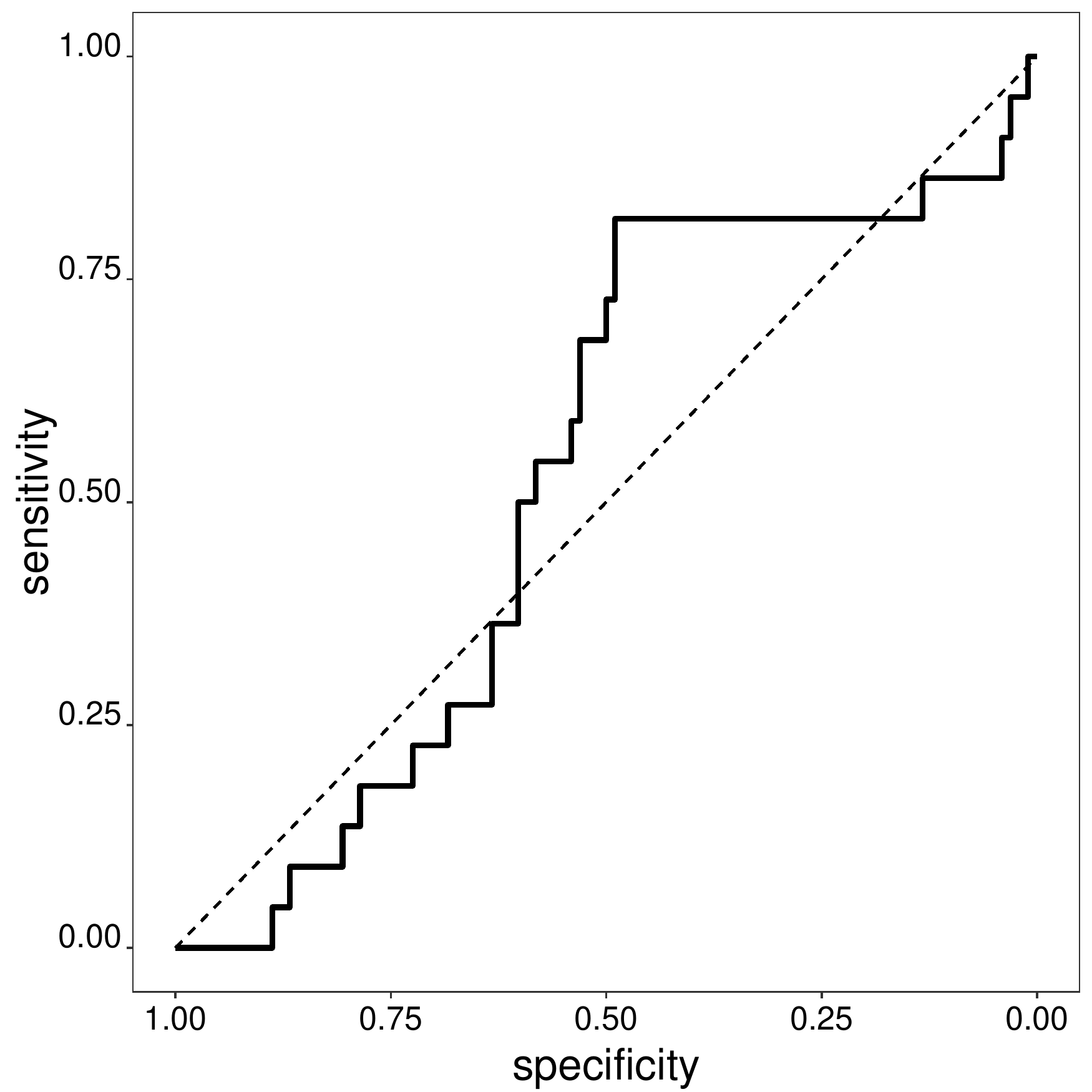}
			\caption{CEA}    
			\label{fig:roc-CEA}
		\end{subfigure}
		\begin{subfigure}[b]{0.24\textwidth}   
			\centering 
			\includegraphics[width=\textwidth]{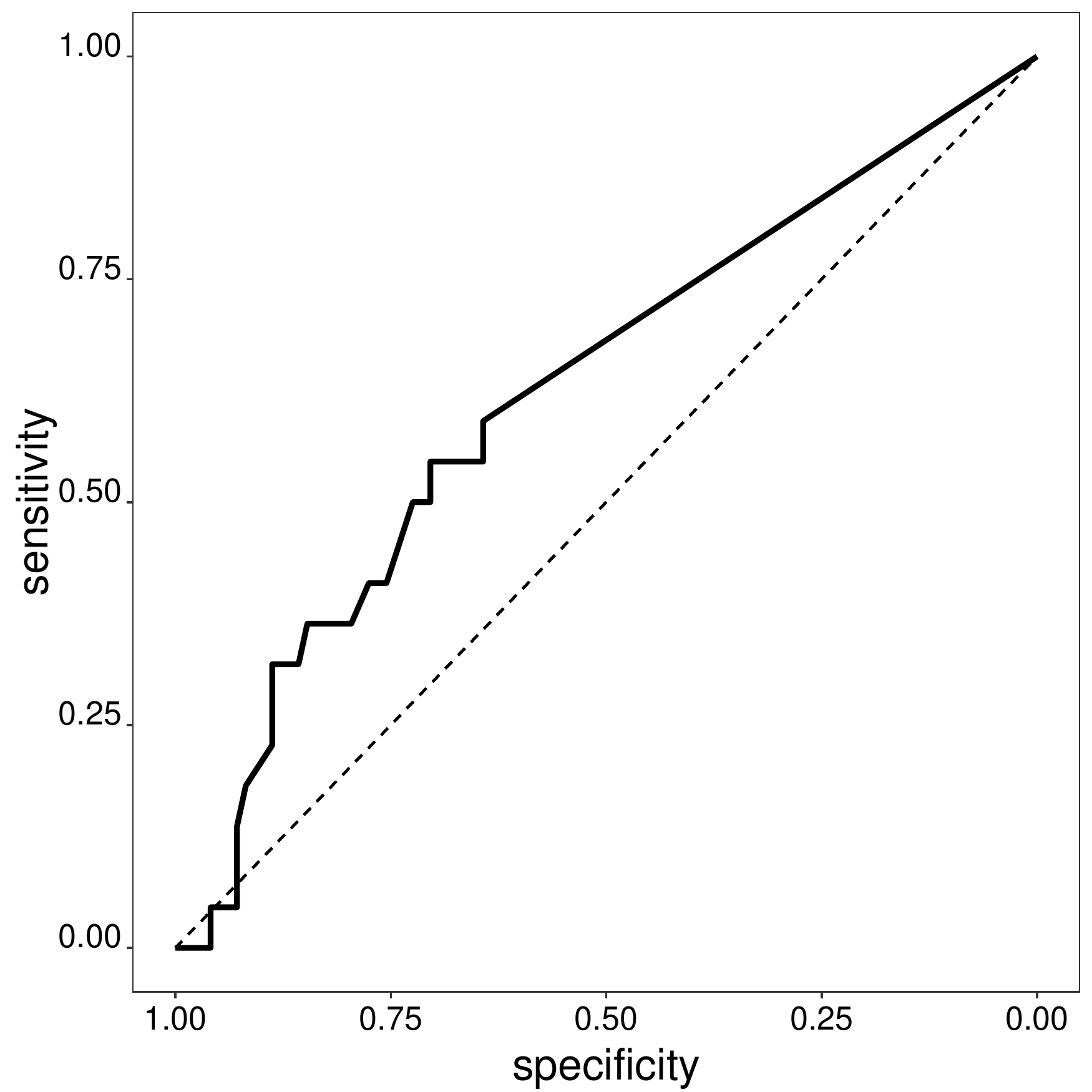}
			\caption{Beta.HCG}    
			\label{fig:roc-BetaHCG}
		\end{subfigure}
		\begin{subfigure}[b]{0.24\textwidth}   
			\centering 
			\includegraphics[width=\textwidth]{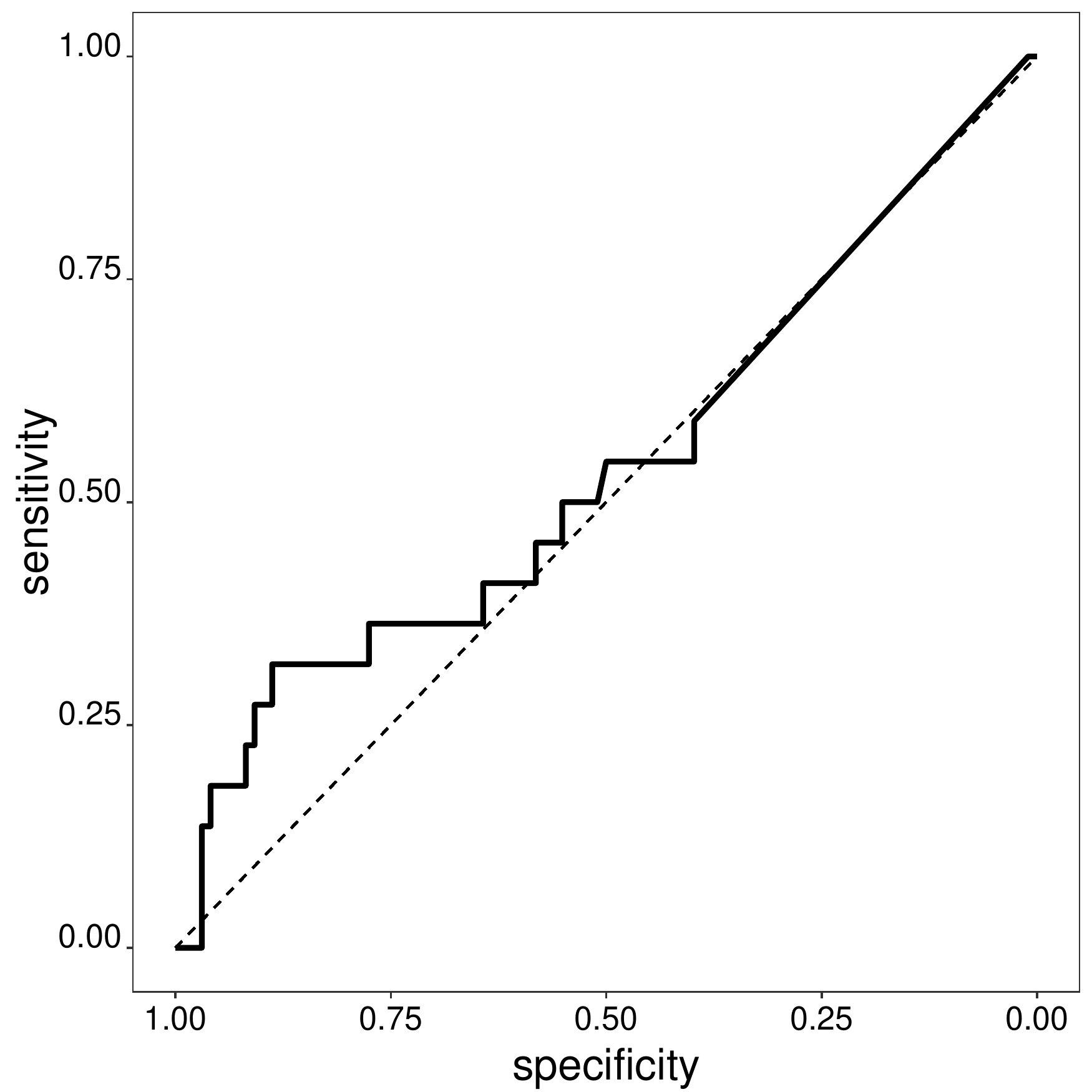}
			\caption{IGFBP.2}    
			\label{fig:roc-IGFBP2}
		\end{subfigure}
		\caption{(\subref{fig:HE4_scatter}) HE4 biomarker levels versus NSCLC recurrence and the fitted logistic regression curve; (\subref{fig:HE4_pvalplot}) $p-$values from Chi-square test using different cut-points for HE4, (c)-(f) ROC curves for top 4 biomarkers
            }\label{realdata_HE4}
	\end{figure}

	One of the biomarkers measured in this study is human epididymis secretory protein 4 (HE4), which is a secretory protein known to be a prognostic factor for NSCLC patients \citep{iwahori2012serum,lamy2015serum,lan2016serum}. 
	Figure \ref{fig:HE4_scatter} shows a scatterplot of the measured HE4 levels and the outcome variable of  recurrence status for the 123 patients. A logistic regression based estimated probability of recurrence curve is overlaid on the scatterplot, This estimated curve does not display a significant slope and both logistic and asymmetric complementary log-log link models for association of HE4 levels with NSCLC recurrence yield $p$-values higher than $0.4$ (Table \ref{tab:real_data}).
	Another common prognostic marker for NSCLC is Carcino Embryonic Antigen (CEA)  \citep{grunnet2012carcinoembryonic,dong2016serum}. \citet{shintani2017prognostic} reported that Stage 1 NSCLC patients with high level of CEA have a higher risk of regional or systemic relapse. In Table \ref{tab:real_data} however, both logistic and complementary log-log link models for association of CEA levels with NSCLC recurrence yield $p$-values higher than $0.6$. The $p$-values for marginal association from logistic regression model for some of the other important serum biomarkers are listed in Table \ref{tab:real_data}. In fact, none of the markers 
	meets the ubiquitous $p < 0.05$ threshold.
	Exploration of joint association via forward and backward selection by AIC \citep{venables2002modern} both returned the null model. Penalized variable selection via   
	Elastic Net \citep{zou2005regularization} and Lasso \citep{tibshirani1996regression}, utilizing different recommended choices of the penalty parameters, also yield the null as the selected model. We further explore Surely Independent Screening \citep{fan2010sure}, which iterates between screening based on marginal association and selection by joint association using the `SCAD' penalty. After many iterations, this approach also returns the null model as the selected model.
	%
		

\begin{table}[!htbp]
\caption{Comparison of $p$-values  of marginal biomarker associations. The $p$-values from univariable binary regression using logit and cloglog links, and the corrections by Miller-Siegmund, Altman, Modified Bonferroni and the proposed Permutation test are reported. For each $p$-value, the FDR corrected $p$-value adjustments are provided in the parentheses. \label{tab:real_data}}
\small
\begin{centering}{}
\begin{tabular}{|c|c c c c c c|}
  \hline
Predictor & logistic & cloglog & Permutation & Miller-Siegmund & Altman & Modified Bonferroni \\ 
  \hline
HE4 & 0.47(0.76) & 0.49(0.74) & 0.01(0.08) & 0.08(0.54) & 0.08(0.54) & 0.03(0.32) \\ 
  CEA & 0.68(0.76) & 0.67(0.74) & 0.05(0.16) & 0.32(0.54) & 0.32(0.54) & 0.21(0.5) \\ 
  beta.HCG & 0.41(0.76) & 0.43(0.74) & 0.08(0.16) & 0.39(0.54) & 0.39(0.54) & 0.25(0.5) \\ 
  IGFBP.2 & 0.44(0.76) & 0.47(0.74) & 0.08(0.16) & 0.39(0.54) & 0.39(0.54) & 0.27(0.5) \\ 
  IGFBP.4 & 0.35(0.76) & 0.37(0.74) & 0.08(0.16) & 0.3(0.54) & 0.3(0.54) & 0.22(0.5) \\ 
  TNFRI & 0.64(0.76) & 0.67(0.74) & 0.1(0.16) & 0.36(0.54) & 0.36(0.54) & 0.31(0.5) \\ 
  FGF.1 & 0.67(0.76) & 0.67(0.74) & 0.11(0.16) & 0.67(0.67) & 0.67(0.67) & 0.5(0.5) \\ 
  IGFBP.3 & 0.1(0.76) & 0.09(0.74) & 0.14(0.17) & 0.49(0.54) & 0.49(0.54) & 0.48(0.5) \\ 
  Angiopoietin.2 & 0.99(0.99) & 0.99(0.99) & 0.16(0.17) & 0.46(0.54) & 0.46(0.54) & 0.44(0.5) \\ 
  TNF.alpha & 0.61(0.76) & 0.63(0.74) & 0.17(0.17) & 0.44(0.54) & 0.44(0.54) & 0.42(0.5) \\ 
   \hline
\end{tabular}
\par\end{centering}{\par}
\end{table}

	These results are clearly negative to the objectives and hypotheses of the study. 
	The Hosmer-Lemeshow test \citep{hosmer1980goodness}  for goodness-of-fit of the  marginal logistic model with HE4 however yields a rather small $p-$value of 0.005 suggesting issues with the logistics regression model. The scatter plot in Figure \ref{fig:HE4_scatter} highlights a few large outlying HE4 values that may be influencing the regression model fit. A scatter plot with CEA (not shown) also displays a few outlying values, however, they represent a different set of patients than the outlying values for HE4. A panel of ROC curves for four biomarkers are shown in Figure \ref{fig:roc-HE4}-\ref{fig:roc-BetaHCG}. In contrast to the weak $p$-values from the logistic regression analyses, these ROC curves show strong to moderate sensitivity and specificity  for NSCLC recurrence.
	%

	Statistical analysis with many features is an increasingly common practice. It is tedious to perform model diagnostics  when association with a large number of features are being explored and for this reason, model diagnostics is often overlooked.
	As we have illustrated in sections
	\ref{sec:binary} and \ref{sec:cont}, the proposed maximal permutation test can be robust to outliers and offers a general blackbox method for making a decision about association without necessarily performing such diagnostics. We employed the maximal permutation test here using chi-square test as the underlying test at each cutpoint. For association of NSCLC recurrence with preoperative levels of the HE4 marker, Figure \ref{fig:HE4_pvalplot} shows a plot of $p-$values obtained at different cut points for the original sequence of the data. This process is repeated for the permuted sequences to obtain the permutation distribution of the test statistic. For comparison, we also report $p$-values based on \citet{miller1982maximally,altman1994danger} and  modified Bonferroni \citep{lausen1996evaluating} approaches. Note that the  \citet{altman1994danger} adjustment is known to be similar to the \citet{miller1982maximally} adjustment for larger $p$-values. As we noted in our simulation studies, these adjustments are often overly conservative and have less power. For association with NSCLC recurrence, the proposed maximal permutation  test reports a $p$-value of $0.008$ for human epididymis secretory protein 4 (HE4) and $p$-value of $0.05$ for Carcino Embryonic Antigen (CEA). 
	respectively.
	After adjusting for multiplicity by the \cite{benjamini1995} False Discovery Rate (FDR) approach, these $p$-values are respectively $0.08$ and $0.16$ but the multiplicity adjustment maintains the ordering of the $p$-values, and  biomarkers HE4 and CEA  still remain at the top among 10 markers ranked by adjusted $p-$value. 

	\section{Concluding Remarks \label{sec:7}}
	The question of association is of prime importance in modern scientific research. 
	Categorization of a predictor is a  popular practice because of its easy interpretability and probably also because we, as humans, process binary information more effectively.
	From statistical point of view, thresholding a predictor has its own advantages as it provides flexibility from parametric models and robustness from outliers. However, systematic search of optimal threshold can strikingly inflate type-1 error.  

	In this article,  we propose a test for association that is free of functional form and distributional assumptions. We illustrate the strong performance of this test as a black box tool for detecting association in diverse settings.
	We provide an extensive set of simulation studies and real data analysis to establish the strong performance of the proposed test. In addition, we propose an innovative application of the proposed methodology in feature screening under sparsity. 
 \vspace*{0.1in}  
 
\noindent \underline{Acknowledgement}
The computational work in this manuscript used resources of the Center for Research Computing and Data at Northern Illinois University.
 \vspace*{-0.1in} 

\bibliographystyle{apa}
\baselineskip = 10pt
\bibliography{permutation}

\begin{thebibliography}{}

\bibitem[\protect\astroncite{Aberle~DR}{2011}]{national2011reduced}
Aberle~DR, Adams~AM, B. C. B. W. C. J. F. R. G. I. G. C. M. P. S.~J. (2011).
\newblock Reduced lung-cancer mortality with low-dose computed tomographic
  screening: The national lung screening trial research team.
\newblock {\em New England Journal of Medicine}, 365(5):395--409.

\bibitem[\protect\astroncite{Altman et~al.}{1994}]{altman1994danger}
Altman, D.~G., Lausen, B., Sauerbrei, W., and Schumacher, M. (1994).
\newblock Dangers of using “optimal” cutpoints in the evaluation of
  prognostic factors.
\newblock {\em JNCI: Journal of the National Cancer Institute},
  86(11):829--835.

\bibitem[\protect\astroncite{Benjamini and Hochberg}{1995}]{benjamini1995}
Benjamini, Y. and Hochberg, Y. (1995).
\newblock Controlling the false discovery rate: a practical and powerful
  approach to multiple testing.
\newblock {\em Journal of the Royal statistical society: series B
  (Methodological)}, 57(1):289--300.

\bibitem[\protect\astroncite{Breiman}{2001}]{breiman2001random}
Breiman, L. (2001).
\newblock Random forests.
\newblock {\em Machine learning}, 45(1):5--32.

\bibitem[\protect\astroncite{Breiman et~al.}{1984}]{breiman1984classification}
Breiman, L., Friedman, J., Stone, C.~J., and Olshen, R.~A. (1984).
\newblock {\em Classification and regression trees}.
\newblock CRC press.

\bibitem[\protect\astroncite{Charloux et~al.}{1997}]{charloux1997prognostic}
Charloux, A., Hedelin, G., Dietemann, A., Ifoundza, T., Roeslin, N., Pauli, G.,
  and Quoix, E. (1997).
\newblock Prognostic value of histology in patients with non-small cell lung
  cancer.
\newblock {\em Lung Cancer}, 17(1):123--134.

\bibitem[\protect\astroncite{Chung and Romano}{2013}]{chung2013}
Chung, E. and Romano, J.~P. (2013).
\newblock Exact and asymptotically robust permutation tests.
\newblock {\em Ann. Statist.}, 41(2):484--507.

\bibitem[\protect\astroncite{Dong et~al.}{2016}]{dong2016serum}
Dong, Y., Zheng, X., Yang, Z., Sun, M., Zhang, G., An, X., Pan, L., Zhang, S.,
  et~al. (2016).
\newblock Serum carcinoembryonic antigen, neuron-specific enolase as biomarkers
  for diagnosis of nonsmall cell lung cancer.
\newblock {\em Journal of cancer research and therapeutics}, 12(5):34.

\bibitem[\protect\astroncite{Doria-Rose and Szabo}{2010}]{doria2010screening}
Doria-Rose, V.~P. and Szabo, E. (2010).
\newblock Screening and prevention of lung cancer.
\newblock {\em Lung cancer: a multidisciplinary approach to diagnosis and
  management}, pages 53--72.

\bibitem[\protect\astroncite{Dudoit and Van
  Der~Laan}{2007}]{dudoit2007multiple}
Dudoit, S. and Van Der~Laan, M.~J. (2007).
\newblock {\em Multiple testing procedures with applications to genomics}.
\newblock Springer Science \& Business Media.

\bibitem[\protect\astroncite{Fan et~al.}{2011}]{fan2011nonparametric}
Fan, J., Feng, Y., and Song, R. (2011).
\newblock Nonparametric independence screening in sparse ultra-high-dimensional
  additive models.
\newblock {\em Journal of the American Statistical Association},
  106(494):544--557.

\bibitem[\protect\astroncite{Fan and Lv}{2008}]{fan2008sure}
Fan, J. and Lv, J. (2008).
\newblock Sure independence screening for ultrahigh dimensional feature space.
\newblock {\em Journal of the Royal Statistical Society: Series B (Statistical
  Methodology)}, 70(5):849--911.

\bibitem[\protect\astroncite{Fan et~al.}{2009}]{Fan2009}
Fan, J., Samworth, R., and Wu, Y. (2009).
\newblock Ultrahigh dimensional feature selection: Beyond the linear model.
\newblock {\em J. Mach. Learn. Res.}, 10:2013--2038.

\bibitem[\protect\astroncite{Fan et~al.}{2010}]{fan2010sure}
Fan, J., Song, R., et~al. (2010).
\newblock Sure independence screening in generalized linear models with
  np-dimensionality.
\newblock {\em The Annals of Statistics}, 38(6):3567--3604.

\bibitem[\protect\astroncite{Fisher}{1936}]{fisher1936design}
Fisher, R.~A. (1936).
\newblock Design of experiments.
\newblock {\em British Medical Journal}, 1(3923):554.

\bibitem[\protect\astroncite{Good}{2013}]{good2013permutation}
Good, P. (2013).
\newblock {\em Permutation tests: a practical guide to resampling methods for
  testing hypotheses}.
\newblock Springer Science \& Business Media.

\bibitem[\protect\astroncite{Grunnet and
  Sorensen}{2012}]{grunnet2012carcinoembryonic}
Grunnet, M. and Sorensen, J. (2012).
\newblock Carcinoembryonic antigen (cea) as tumor marker in lung cancer.
\newblock {\em Lung cancer}, 76(2):138--143.

\bibitem[\protect\astroncite{Halpern}{1982}]{halpern1982maximally}
Halpern, J. (1982).
\newblock Maximally selected chi square statistics for small samples.
\newblock {\em Biometrics}, pages 1017--1023.

\bibitem[\protect\astroncite{Hochberg and
  Tamhane}{1987}]{Hochberg:1987:MCP:39892}
Hochberg, Y. and Tamhane, A.~C. (1987).
\newblock {\em Multiple Comparison Procedures}.
\newblock John Wiley \& Sons, Inc., New York, NY, USA.

\bibitem[\protect\astroncite{Hoeffding}{1952}]{hoeffding1952large}
Hoeffding, W. (1952).
\newblock The large-sample power of tests based on permutations of
  observations.
\newblock {\em The Annals of Mathematical Statistics}, pages 169--192.

\bibitem[\protect\astroncite{Hosmer and Lemesbow}{1980}]{hosmer1980goodness}
Hosmer, D.~W. and Lemesbow, S. (1980).
\newblock Goodness of fit tests for the multiple logistic regression model.
\newblock {\em Communications in statistics-Theory and Methods},
  9(10):1043--1069.

\bibitem[\protect\astroncite{Iwahori et~al.}{2012}]{iwahori2012serum}
Iwahori, K., Suzuki, H., Kishi, Y., Fujii, Y., Uehara, R., Okamoto, N.,
  Kobayashi, M., Hirashima, T., Kawase, I., and Naka, T. (2012).
\newblock Serum he4 as a diagnostic and prognostic marker for lung cancer.
\newblock {\em Tumor Biology}, 33(4):1141--1149.

\bibitem[\protect\astroncite{Janssen}{1997}]{JANSSEN19979}
Janssen, A. (1997).
\newblock Studentized permutation tests for non-i.i.d. hypotheses and the
  generalized behrens-fisher problem.
\newblock {\em Statistics \& Probability Letters}, 36(1):9 -- 21.

\bibitem[\protect\astroncite{Janssen et~al.}{2003}]{janssen2003bootstrap}
Janssen, A., Pauls, T., et~al. (2003).
\newblock How do bootstrap and permutation tests work?
\newblock {\em The Annals of Statistics}, 31(3):768--806.

\bibitem[\protect\astroncite{Kanodra et~al.}{2015}]{kanodra2015screening}
Kanodra, N.~M., Silvestri, G.~A., and Tanner, N.~T. (2015).
\newblock Screening and early detection efforts in lung cancer.
\newblock {\em Cancer}, 121(9):1347--1356.

\bibitem[\protect\astroncite{Lamy et~al.}{2015}]{lamy2015serum}
Lamy, P.-J., Plassot, C., and Pujol, J.-L. (2015).
\newblock Serum he4: an independent prognostic factor in non-small cell lung
  cancer.
\newblock {\em PloS one}, 10(6):e0128836.

\bibitem[\protect\astroncite{Lan et~al.}{2016}]{lan2016serum}
Lan, W.-G., Hao, Y.-Z., Xu, D.-H., Wang, P., Zhou, Y.-L., and Ma, L.-B. (2016).
\newblock Serum human epididymis protein 4 is associated with the treatment
  response of concurrent chemoradiotherapy and prognosis in patients with
  locally advanced non-small cell lung cancer.
\newblock {\em Clinical and Translational Oncology}, 18(4):375--380.

\bibitem[\protect\astroncite{Lausen and Schumacher}{1992}]{lausen1992maximally}
Lausen, B. and Schumacher, M. (1992).
\newblock Maximally selected rank statistics.
\newblock {\em Biometrics}, pages 73--85.

\bibitem[\protect\astroncite{Lausen and
  Schumacher}{1996}]{lausen1996evaluating}
Lausen, B. and Schumacher, M. (1996).
\newblock Evaluating the effect of optimized cutoff values in the assessment of
  prognostic factors.
\newblock {\em Computational Statistics \& Data Analysis}, 21(3):307--326.

\bibitem[\protect\astroncite{Lehmann et~al.}{1949}]{lehmann1949theory}
Lehmann, E.~L., Stein, C., et~al. (1949).
\newblock On the theory of some non-parametric hypotheses.
\newblock {\em The Annals of Mathematical Statistics}, 20(1):28--45.

\bibitem[\protect\astroncite{Li et~al.}{2012}]{Li2012}
Li, R., Zhong, W., and Zhu, L. (2012).
\newblock Feature screening via distance correlation learning.
\newblock {\em Journal of the American Statistical Association},
  107(499):1129--1139.
\newblock PMID: 25249709.

\bibitem[\protect\astroncite{Mazumdar and
  Glassman}{2000}]{mazumdar2000categorizing}
Mazumdar, M. and Glassman, J.~R. (2000).
\newblock Categorizing a prognostic variable: review of methods, code for easy
  implementation and applications to decision-making about cancer treatments.
\newblock {\em Statistics in medicine}, 19(1):113--132.

\bibitem[\protect\astroncite{Meier et~al.}{2009}]{meier2009high}
Meier, L., Van~de Geer, S., B{\"u}hlmann, P., et~al. (2009).
\newblock High-dimensional additive modeling.
\newblock {\em The Annals of Statistics}, 37(6B):3779--3821.

\bibitem[\protect\astroncite{Miller and Siegmund}{1982}]{miller1982maximally}
Miller, R. and Siegmund, D. (1982).
\newblock Maximally selected chi square statistics.
\newblock {\em Biometrics}, 38(4):1011--1016.

\bibitem[\protect\astroncite{Mulshine and D'Amico}{2014}]{mulshine2014issues}
Mulshine, J.~L. and D'Amico, T.~A. (2014).
\newblock Issues with implementing a high-quality lung cancer screening
  program.
\newblock {\em CA: a cancer journal for clinicians}, 64(5):351--363.

\bibitem[\protect\astroncite{Pitman}{1937}]{pitman1937significance}
Pitman, E.~J. (1937).
\newblock Significance tests which may be applied to samples from any
  populations.
\newblock {\em Supplement to the Journal of the Royal Statistical Society},
  4(1):119--130.

\bibitem[\protect\astroncite{Shintani et~al.}{2017}]{shintani2017prognostic}
Shintani, T., Matsuo, Y., Iizuka, Y., Mitsuyoshi, T., Mizowaki, T., and
  Hiraoka, M. (2017).
\newblock Prognostic significance of serum cea for non-small cell lung cancer
  patients receiving stereotactic body radiotherapy.
\newblock {\em Anticancer research}, 37(9):5161--5167.

\bibitem[\protect\astroncite{Székely and Rizzo}{2009}]{distcorr2009}
Székely, G.~J. and Rizzo, M.~L. (2009).
\newblock Brownian distance covariance.
\newblock {\em Ann. Appl. Stat.}, 3(4):1236--1265.

\bibitem[\protect\astroncite{Székely et~al.}{2007}]{distcorr2007}
Székely, G.~J., Rizzo, M.~L., and Bakirov, N.~K. (2007).
\newblock Measuring and testing dependence by correlation of distances.
\newblock {\em Ann. Statist.}, 35(6):2769--2794.

\bibitem[\protect\astroncite{Tibshirani}{1996}]{tibshirani1996regression}
Tibshirani, R. (1996).
\newblock Regression shrinkage and selection via the lasso.
\newblock {\em Journal of the Royal Statistical Society. Series B
  (Methodological)}, pages 267--288.

\bibitem[\protect\astroncite{Tukey}{1960}]{tukey1960survey}
Tukey, J.~W. (1960).
\newblock A survey of sampling from contaminated distributions.
\newblock {\em Contributions to probability and statistics}, pages 448--485.

\bibitem[\protect\astroncite{Venables and Ripley}{2002}]{venables2002modern}
Venables, V. and Ripley, B. (2002).
\newblock Modern applied statistics with s. chambers sj, eddy w, hardle w,
  sheater s, tierney l.

\bibitem[\protect\astroncite{Wald and Wolfowitz}{1944}]{wald1944statistical}
Wald, A. and Wolfowitz, J. (1944).
\newblock Statistical tests based on permutations of the observations.
\newblock {\em The Annals of Mathematical Statistics}, 15(4):358--372.

\bibitem[\protect\astroncite{Wang}{2009}]{Wang2009}
Wang, H. (2009).
\newblock Forward regression for ultra-high dimensional variable screening.
\newblock {\em Journal of the American Statistical Association},
  104(488):1512--1524.

\bibitem[\protect\astroncite{Williams et~al.}{1981}]{williams1981survival}
Williams, D., Pairolero, P., Davis, C., Bernatz, P., Payne, W., Taylor, W.,
  Uhlenhopp, M., and Fontana, R. (1981).
\newblock Survival of patients surgically treated for stage i lung cancer.
\newblock {\em The Journal of thoracic and cardiovascular surgery},
  82(1):70--76.

\bibitem[\protect\astroncite{Yu and Moyeed}{2001}]{yu2001bayesian}
Yu, K. and Moyeed, R.~A. (2001).
\newblock Bayesian quantile regression.
\newblock {\em Statistics \& Probability Letters}, 54(4):437--447.

\bibitem[\protect\astroncite{Zeileis}{2006}]{zeileis2006object}
Zeileis, A. (2006).
\newblock Object-oriented computation of sandwich estimators.
\newblock {\em Journal of Statistical Software, Articles}, 16(9):1--16.

\bibitem[\protect\astroncite{Zou and Hastie}{2005}]{zou2005regularization}
Zou, H. and Hastie, T. (2005).
\newblock Regularization and variable selection via the elastic net.
\newblock {\em Journal of the Royal Statistical Society: Series B (Statistical
  Methodology)}, 67(2):301--320.

\end{thebibliography}
	\newpage

\end{document}